# Organizational Practices and Socio-Technical Design of Human-Centered AI

Thomas Herrmann, Institute of Work Science, University of Bochum, Bochum, Germany, Thomas.herrmann@rub.de, https://orcid.org/0000-0002-9270-4501

**Abstract:**

This contribution explores how the integration of Artificial Intelligence (AI) into organizational practices can be effectively framed through a socio-technical perspective to comply with the requirements of Human-centered AI (HCAI). Instead of viewing AI merely as a technical tool, the analysis emphasizes the importance of embedding AI into communication, collaboration, and decision-making processes within organizations from a human-centered perspective. Ten case-based patterns illustrate how AI support of predictive maintenance can be organized to address quality assurance and continuous improvement and to provide different types of support for HCAI. The analysis shows that AI adoption often requires and enables new forms of organizational learning, where specialists jointly interpret AI output, adapt workflows, and refine rules for system improvement. Different dimensions and levels of socio-technical integration of AI are considered to reflect the effort and benefits of keeping the organization in the loop.

## 1. Introduction

### 1.1. Human-Centered AI and Socio-Technical Design

The basic perspective of this chapter is that human-centered AI (HCAI) is built on socio-technical design and that not only technical aspects, but also organizational practices are relevant for achieving the goals of HCAI.

HCAI is understood as an approach that maintains the relevance of human contributions under conditions of increasing automation of labor by keeping the human in the loop (Crootof et al., 2023). HCAI can be characterized by the following goals:

- Empowering people by supporting their autonomy and creativity (Shneiderman, 2022),
- Opening up opportunities for humans to use their experience, knowledge and skills, and to further develop their own capabilities (Dellermann et al., 2019),
- Supporting humans' interests, needs and values, such as fairness, wellbeing etc.,
- Enabling a kind of AI-supported task handling and decision making where humans are made capable of understanding the generation of AI-output as a basis to influence and control AI behavior (Xu & Gao, 2025),
- Offering the experience of being augmented and in the lead when interacting with AI (Xu & Gao, 2025).

These goals can be further detailed by referring to principles of HCAI design and socio-technical design (Section 2.1). To meet these goals, systematic HCAI-frameworks have been developed (Shneiderman, 2022; Xu & Gao, 2025) that cover the whole cycle of AI design, development, implementation, use and continuous appropriation and adaptation.

To achieve the goals of HCAI, it is relevant to employ the insights of socio-technical analysis and design (Cherns, 1976; Mumford, 2000; Trist, 1981). Socio-technical design has the following characteristics: Humans usually do not work in isolation but together with others—mostly in teams. Thus, the organization of collaborative work, the various roles of human actors can play, and the regulation of how these roles are coordinated are of indispensable relevance. Consequently, socio-technical design seeks to integrate three aspects for the processes of task execution within work systems: human needs and capabilities, organizational structures and technical artefacts (Herrmann, 2009; Kirsch et al., 1995). Processes of task execution are part of the work system for which the joint optimization of technology, people's qualification and organizational structures are pursued. This optimization seeks to balance human interests and needs with productivity, performance and efficiency. Furthermore, Herrmann et al. (2022) emphasize that the socio-technical intertwinement of organizational practices and technical artefacts is based on human-computer interaction. They consider organizational practices as part of social practices that include collaboration and communication between humans. Additionally, socio-technical systems are never complete (Cherns, 1987) and need ongoing design that includes feedback loops and participation of its users and stakeholders. Evolution and flexibility are crucial facets (Herrmann, et al., 2022) of socio-technical system. Flexibility and the need for continuous evolution of socio-technical systems is grounded in the characteristics of social systems (Luhmann, 1995). A deliberate understanding of socio-technical design (Winby & Xu, 2026) is of crucial relevance for HCAI.

Organizational design is understood as a strategic process that specifies how work is divided, the duties of roles, workflows of collaboration, communication channels, and possibilities for flexibility and agility (Holbeche, 2019). Organizational design can partially shape and influence organizational practices but cannot fully determine them (see Section 2.2). Taking organizational design into account focuses on the meso-level of social analysis. This level can be understood as lying in-between the micro level and the macro level. The micro level mainly considers individual workspaces for task execution and their conditions, such as structures of human computer interaction, available technical functionality to carry out a task, human skills to complement the task etc. The macro level addresses the conditions established by society as a whole. Typically, the laws, socio-cultural norms and rituals, political structures, economic conditions etc. Rahwan (2018) addresses the macro level and points out the relevance of keeping the society in the loop when designing and applying AI. The idea here is that the technical structure and features of AI should mirror a society's ethical values and regulations. By contrast, from the perspective of HCAI, it is mainly the micro level that is addressed, e.g. by emphasizing the role of AI as a (powerful)tool (Shneiderman, 2020) or by focusing interaction modes (Herrmann, 2022).

The meso-level is mostly not systematically considered in research on HCAI. An exception is the intelligent socio-technical systems framework of XU and Gao (2025) which proposes a four-level hierarchy including organization and ecosystems on the meso-level. They emphasize that exclusively focusing on single organizations is no longer appropriate from a socio-technical point of view, since human-AI systems can also be interconnected in cooperation between organizations, communities, social networks etc.

To further elaborate the relevance of the meso-level, this chapter emphasizes that the goals of HCAI can only be met if appropriate modes of human-computer interaction are combined with accompanying organizational design that support them by keeping the organization in the loop (Herrmann & Pfeiffer, 2023a).

## 1.2. Objectives and Scope

The objective of this chapter is to address the gap that the socio-technical aspects of organizational design and practices are not sufficiently considered and detailed in the HCAI discourse. Therefore, the leading questions for this chapter are:

1. Which existing approaches that address the relationship between organizational design and information technology are relevant for HCAI?
2. What are different ways of organizationally embedding the usage of AI and how do these ways influence the goals of HCAI?
3. Which dimensions help understand the organizational effort implied by various patterns of embedding AI, and what are HCAI-related drivers of investing this effort?

Answering these questions is important for establishing organizational practices that accompany the implementation and usage of AI on the meso-level. This involves a range of disciplines: In computer science the discipline of computer-supported cooperative work (CSCW) is relevant since it helps to understand the needs of people collaborating with others, how collaborative AI should behave, how awareness between actors is managed and how coordination issues are solved etc. Not only CSCW but also Information Systems, in particular business process management, deal with the question of how to technically support and coordinate workflow management. Socio-technical design traditionally takes the perspective of teamwork into account and places early emphasis on the inherent incompleteness of socio-technical systems. Furthermore, legal and ethical considerations are relevant for avoiding biases and maintaining privacy. Work design and industrial sociology help to understand requirements for task structuring and task sharing while philosophy and anthropology address the differences between human and AI.

Analyzing and designing organizational practices is a truly multi-disciplinary challenge. While the domain of human-computer interaction already involves technical, psychological, and medical knowledge, the relevance of different research areas becomes even broader when the meso- and macro-level are additionally considered. This also indicates that designing appropriate organizational practices and their interplay with AI is not a design task to be completed in one step but requires constant evaluation and iteration to progress from one level of maturity to the next.

To answer the first question, the background (Section 2) refers to approaches that reflect the quality of work design, describe the emergence of organizational practices and organizational culture, relates them to organizational learning, differentiates management strategies, relates business process management to AI and discusses the relevance of teamwork and networking structures. For each approach, its relevance for HCAI is outlined. To answer the second question, two case studies are introduced (Section 3) that substantiate the concept of keeping the organization in the loop. Based on the case studies, ten different patterns of how organizations can manage the implementation and integration of AI are derived (Section 4). The patterns demonstrate that organizations can choose between different ways of applying AI. These patterns make it apparent that the possible ways of organizationally embedding AI have different effects on achieving the goals of HCAI. The socio-technical implementation of HCAI will not occur automatically. Stakeholders involved, especially managers, need to be convinced that the additional effort required for deliberate organizational design is a worthwhile investment in order to achieve HCAI goals. Therefore, eight different dimensions of organizational design are derived from the patterns and discussed with respect to organizational effort (Section 5). Depending on this effort, different levels of HCAI related benefits can be achieved. The Conclusion (Section 6) emphasizes the necessity of keeping the organization in the loop and addresses contradictions that must be managed. Consequently, there is no single best solution for each case of organizationally implementing AI. Investing effort in a socio-technically informed implementation of HCAI that keeps the organization in the loop appears more beneficial the more a company strives for continuous improvement and innovation. Thus, this chapter is not a case study about best practices but about possible variations. It considers the socio-technical integration of AI as a subject of continuous improvement and evolution.

# 2. Background – Approaches of Addressing Organizational Practices in the Context of IT Usage

This section analyzes and discusses the implementation and usage of AI in light of various concepts and characteristics of organizational design. Research can consider how applying these concepts influences the outcomes of AI usage and helps organizations become resilient against failures and risks, as well improve and innovate. Management needs to consider these concepts to evaluate their relevance for implementing AI in concrete settings.

The concepts presented in this section are:

- Human-centered work design
- Organizational practices and organizational culture
- Organizational learning and change management
- Management strategies
- Teamwork
- Networking and communities
- Business Process Management

It is not possible to present these concepts in detail but at least to demonstrate that they still have relevance to the question of how AI is implemented and used. Therefore, these approaches are briefly described and can be referred to later when analyzing the case studies that are introduced below.

## 2.1. Human-Centered Work Design

Scholars in the domain of work design (Grote et al., 2000; Hackman & Oldham, 1980; Mumford, 1983) propose design principles that pursue a balance between efficiency and productivity on the one side and human needs and interests in well-being, personal development, etc., on the other side. This balance addresses the design of work systems that are at the core of socio-technical design. "A work system is a system in which human participants and/or machines perform business processes using information, technologies, and other resources to produce products and/or services for internal or external customers. (Alter, 2002, p. 92)" To pursue this balance means asking for the appropriate division of labor between humans and machines. An appropriate division does not just aim to leave some tasks to humans. By contrast, it requires a mix of challenges that employ human strength and capabilities and support their development. Well-designed tasks as part of a socio-technical work system need to be complete, including preparation, planning, execution and control, and to be perceivable as a whole piece of work that has a significant impact on others. A human-centered approach means that human needs must be given priority when deciding how labor should be divided between humans and machines. (Cooley, 1996).

Work design is closely intertwined with organizational and socio-technical design. Herrmann et al. incorporate principles of work design in a series of eight heuristics that help evaluate the quality of socio-technical systems (Herrmann, et al., 2022). They require that tasks involve a variety of different characteristics, particularly with regard to the use of different skills and knowledge, so that people can experience that greater effort during work can lead to greater benefits. The underlying concept is job enrichment (Hackman et al., 1975). The way sub-tasks are designed and allocated to either humans or machines must allow for autonomy—both for groups and individuals—in decision-making, flexible use of tools, coordination with others, etc. This flexibility is a prerequisite for the continuous evolution of the socio-technical system, which is driven by workforce participation and efforts to appropriate the technology employed. Appropriation includes negotiating the way of using and tailoring technology as well as going beyond the intentions of the designers and the rules of the organization initially associated with the artifact being introduced (Pipek, 2005). To be able to exercise autonomy and oversight, transparency and visibility are needed so that workers can understand what they and others have already achieved, what future activities are possible or required, and which methods and tools are available. In this context, it is acknowledged that people need appropriate feedback about the effectiveness of their performance. A further need is that tasks allow workers to interact with others and include opportunities and requirements to communicate with other people. Finally, well-structured tasks require employees to use the full range of their skills and offer them opportunities for learning and personal development. This requirement aims to improve compatibility between the way of how staff are prepared on the one side and the tasks to be completed, technologies to be used and customers' expectations to be met on the other side (Herrmann, et al., 2022).

**Relevance for human-centered AI**

The work design principles provide well-established criteria for what is important for the human-centered implementation and usage of AI. They are also mirrored by socio-technical design principles for HCAI as provided by Xu and Gao (2025). These principles emphasize joint optimization, human-led collaboration and decision making with AI, as well as human-AI complementarity and co-evolution. They go beyond the conventional socio-technical principles by adding the aspect of ethical and responsible design that addresses fairness, avoiding biases and guaranteeing privacy.

Joint optimization and human-AI complementarity focus on the question of how sub-tasks are allocated and shared between humans and AI. It should be noted that transferring tasks from humans to AI can not only reduce human activities but also lead to new sub-tasks that extend the scope of human contributions (Muller & Weisz, 2022). This can result from task design guided by the principles mentioned above. It must be avoided that only marginal tasks remain on the human side. Control and oversight over AI are important aspects of autonomy, not only to mitigate risks but also as a psychological need. Control includes the flexibility to decide how AI output is further processed. What is new with AI is that its socio-technical interplay with socio-technical task handling is not only a subject of design principles and guidelines, but can also help fulfill these requirements (Herrmann, 2024).

## 2.2. Organizations and Organizational Practices – from Rules to Conventions

From the viewpoint of this chapter, organizations—be they commercial enterprises or public institutions—have the following characteristics: They are subject to their own logic, integrated into complex external environments, and divided internally by competition among departments with divergent interests. Decisions, whether technically derived or generated by humans, need to be negotiated and processed within the organization (Herrmann & Pfeiffer, 2023a; Phillips & Lawrence, 2012). Organizational regulations, as issued by the management, are only implemented as far as they find their way into everyday practices. The should-be model of an organization, as defined in the regulations issued by management, is only partially mirrored in everyday organizational practices.

Organizational practices are characterized by routinized collective activities and processes that have evolved and are possibly under continuous development to achieve the objectives of an organization and to maintain its culture. Organizational practices are based on formal and informal roles, procedures and communication (Aubouin-Bonnaventure et al., 2021; Verbeke, 2000). They are less based on formal regulations but on conventions that develop gradually (Mark, 2002). Conventions are recurring patterns of behavior, for which a group expects its members to comply. This compliance is based on the experience that the conventions are suitable to solve coordination problems to the benefit of the group or the entire organization.

The way in which management regulations are transformed into conventions depends decisively on the type of organizational culture that characterizes a company. Organizational culture focuses on a wide range of aspects, such as values, beliefs, norms, and attitudes, that are shared by the members of an organization and guide their behavior (Gutterman, 2024; Schein, 1990; Tadesse Bogale & Debela, 2024). On the one hand, organizational culture helps develop and maintain organizational practices and their underlying conventions. On the other hand, organizational culture is also an outcome of organizational practices. Organizational culture is less the result of regulations but of communicational routines. These routines include formal as well as informal types of communication. For the development of organizational culture, informal communication has a decisive influence. Kraut et al. (1993) characterize informal communication by referring to the absence of specifications such as the place and time of communication, its goals, and participants. Informal communication is usually not directly task-related and its content cannot be directly attributed to a specific person in responsibility. Its understanding depends decisively on common ground and shared context (Kienle & Herrmann, 2003). Organizational practices, organizational culture and informal communication can hardly be specified by top-down regulations. It turns out that the social side of socio-technical systems cannot be deterministically designed and thus is different compared with conventional technology (Fischer & Herrmann, 2011). Consequently, the design of organizational practices has to deal with the tension between formal and informal structures and behavior (Kreutzer et al., 2016).

**Relevance for human-centered AI:**

How AI is used within organizational practices and how far this usage complies with ethical guidelines and socio-technical principles also depends on a mixture of formal and informal structures. Consequently, the way AI is appropriated is not strictly manageable by top down regulations. One has to be aware that how organizational practices using AI develop is a subtle process of communication negotiation, and feedback loops. They drive the evolution of conventions about which kinds of employing AI comply with people's expectations and interests. This becomes even more relevant if AI is implemented as an agent that can take part in these negotiations. The potential of AI to adapt itself also limits the possibilities for deterministic top-down regulation and can cause unsolicited variances. However, these variances can possibly be a source of innovation or improvement, and of varieties that enrich human work.

If one accepts the unavoidable impreciseness and incompleteness (Goedicke & Herrmann, 2008; Herrmann & Loser, 1999) of socio-technical systems, it is also acceptable that organizational design has to find ways of how to deal with them. This is also applicable for dealing with potential imperfections and insufficiencies in the context of HCAI. Organizational strategies in this context can point toward two different directions: Organizations can restrict their AI-usage to those systems that have proven high reliability and correctness, or they can also use less reliable versions that help increase efficiency but require strict human oversight and autonomy in accordance with HCAI. The second option is a vision of socio-technical design where social practices compensate for the weaknesses of AI and vice versa. In such a context, humans can maintain an indispensable role where they can still employ their knowledge and skills and further develop them. In these circumstances, it is important to calibrate workers' trust in AI (Okamura & Yamada, 2020), as under-trust can lead to inefficiency, and over-trust can be a source of serious risks that can cause harm to people or equipment.

## 2.3. Organizational Learning, Knowledge Management and Change Management

The continuous evolution of socio-technical systems and organizational practices is closely related to processes of learning grounded in ongoing everyday experience during task handling by the workforce. These learning processes result from individual learning activities during task handling. The learning outcomes are not simply added together but are intertwined in such a way that they lead to learning progress at the organizational level. Processes of organizational learning (Argote et al., 2021) begin with individual experiences and are completed through systematic reflection on those experiences (Boud, 1985). Reflection is a source of self-efficacy (Bandura, 1977) that shapes and mirrors employees' beliefs about how they are capable of mastering their tasks and challenges. Reflection includes the search of relevant insights, knowledge creation as well as documentation and retention, and finally knowledge dissemination. Thus, organizational learning is closely related to knowledge management and the implementation of knowledge management systems (Alavi & Leidner, 2001).

Knowledge management as a technical means is not sufficient to drive innovation, improve productivity and effectiveness. It needs to be closely coupled with organizational practices and change management (By, 2005). Only in this way can organizational learning meet the needs of the organization and help provide feedback on how well and efficient changes are taking place. Change management is considered an endeavor of continuous improvement that covers strategies and structures as well as the capabilities needed to meet the dynamic requirements of external and internal customers (Moran & Brightman, 2000). Together with organizational learning, it supports the continuous evolution and improvement of socio-technical systems.

**Relevance for human-centered AI:**

Implementing AI requires understanding and application of change management methods. AI is both a continuous driver and a subject of change and organizational learning. For supporting change and learning, knowledge management systems were considered useful. However, these will be replaced or at least supplemented by AI systems in the future (Jarrahi et al., 2023). Since the functionality and technical characteristics of AI applications will continuously develop in the coming years, organizations' learning and knowledge sharing also need to mirror this development to address HCAI. Implementing and using AI can both be intertwined with change management to make the usage increasingly efficient and to take the needs of the workforce and of customers into account. AI itself – and this is a new dimension – can support organizational learning, knowledge management and change management with regard to the HCAI-oriented socio-technical integration into the organization.

## 2.4. Management Strategies: From Taylorism to Management by Delegation

AI is a promising technology to overcome step-by-step specification of task execution in favor of getting entire tasks done with technology by mirroring communicational procedures of management by delegation and management by objectives. These management strategies (Husainah & Ahmad, 2024; Maciariello, 2016; Piplica et al., 2024) seek to increase efficiency and effectiveness by giving staff sufficient competence and self-confidence to execute tasks as a whole by just specifying goals and constraints. Important conditions to be specified concern the degree of quality that must be achieved. Employees gain motivation and have the option for personal development by making their own decision on how to handle a task and on which tools they will employ. This focus on outcome instead of procedures stands in contrast to Taylorism (Wright, 1993), which is characterized by standardization and centralized control. The prerequisites for task delegation are to make sure that employees are sufficiently skilled to carry out the tasks and to provide them with appropriate feedback to the outcome at the end of and during task execution – especially with respect to the quality to be achieved. Management by delegation only works if employees or teams are granted a certain degree of autonomy.

**Relevance for human-centered AI:**

The performance of conventional IT depends on processes of software-development where the requirements that reflect tasks are engineered and are met by programming as a description of step-by-step procedures. By contrast, tasks can be delegated to AI as a whole. Management styles that are based on management by delegation and management by objectives can be applied in dealing with AI. This might be especially relevant if AI is organizationally integrated as a teammate or collaboration partner (Dubey et al., 2020; O'Neill et al., 2022). However, the delegation metaphor is challenged by the question of how much autonomy can be granted to AI without losing the necessary accountability that must be exercised through human decision-making (Mathebula & Barnard, 2020). Finding ways of integrating human oversight into the delegation of task is especially important for the usage of agentic AI (Acharya et al., 2025) to support HCAI.

## 2.5. Teamwork

Socio-technical design mainly focus on teams (Mumford, 1995) on the social side to foster a perspective that goes beyond the mere consideration of human-machine interaction. Teams can be a subject of organizational design and they develop their specific organizational practices that are also shaped by the technologies applied as described by Trist and Bamforth (1951). Organizational design of teams has to regard their tasks and goals, the number of members, the diversification of their skills and competences, and possibly different personality traits management may want to be combined in a team (Mohrman et al., 1995). To a certain degree, teams can autonomously negotiate how they share tasks and roles between their members. It has to be specified how a team is interacting with its environment, most important who is in charge and allowed to communicate to the outside. The interaction between teams has to be organized and two different teams may partially have the same members. Moving from top-down hierarchies towards teams increases the possibilities for innovation and flexibility and adaptivity to external requirements of customers from inside and outside a company. Functional collaboration and integration of knowledge as well as direct ways of formal and informal communication are success factors of team work (He et al., 2022). The set of members of a team can dynamically change. Deliberate processes of onboarding are crucial for the success of teamwork.

**Relevance for human-centered AI:**

Organizations will have the choice when implementing AI whether only individual handling of certain tasks will change, whether they address a whole team's work or even the entire company or relations with other organizations with whom they collaborate in a network. A team might be an appropriate unit where integration of AI can be tested to evaluate its effects on collaborative work and its coordination. Pursuing this approach requires deliberate consideration of the existing knowledge about characteristics and success factors of teamwork. With respect to HCAI, it has to be decided how far the role of AI is negotiable by the team members and how it affects their autonomy (Berretta et al., 2023). Furthermore, it is a question how far AI will have an onboarding procedure like a new team member (Cai et al., 2019).

## 2.6. Networking and Communities

When introducing new technology, management might attempt to limit its effects on smaller units such as single workspaces or teams. However, technology usually has the potential to also influence the surrounding of these units, and can eventually evolve to affect the whole company as well as parts of the networks with which a company has interdependencies as Orlikowski (1996) demonstrated. Furthermore, individual workers or small enterprises can become members of communities that exchange experience about solving certain problems or using technologies.

Networking organizations—such as inter-organizational alliances or professional communities—seek to realize advantages of decentralized and flexible organizational design, for example, for running complex projects (Powell, 1990). Networks rely on reciprocal interests and trust. They develop and maintain structures with which they can quickly react to changing conditions and dynamic dependencies. Less formal than networks are communities of practice (Wenger, 1999) that support informal knowledge exchange. They develop their own social structures where people support each other and share relevant roles and duties such as helping as facilitators. The requirements for a high level of adaptivity and responsiveness within networks and communities can lead to the dissolvement of fixed boundaries between organizations. This dissolvement can result in renewed organizational practices that foster efficiency and innovation (Gulati et al., 2012). Apparently, interorganizational activities include collaboration and coordination that are mirrored by organizational practices. Consequently, inner-organizational task handling has to be flexibly intertwined with networking relations (Castañer & Oliveira, 2020).

**Relevance for human-centered AI:**

Human-centered implementation and running of AI applications involve several kinds of specialists who take care of technical maintenance, reconfiguration, data provision and data curation. These specialists are mostly not available in small or medium enterprises and have to be involved via networking. Successful AI-usage will although include other organizations, for example, customers or suppliers. Changes in AI-equipment and AI-usage of one company will probably influence their exchange with other companies. Vice versa, changes in the context of an enterprise will have to be reacted on—also with the help of AI and the way of how AI is used—to maintain human-centeredness.

## 2.7. Business Process Management and Quality Assurance

In cases where routine tasks are collaboratively repeated in a similar way, organizational practices are based on and accompanied by business process management (BPM) and workflow management systems. Business process management (Dumas et al., 2013) has developed a set of methods with which processes can be analyzed and documented on various levels of detail. Referring to the analysis and documentation, proposals for process improvement and quality assurance can be developed based on elaborated best practices (Reijers & Mansar, 2005) that can be utilized as a subset of socio-technical guidelines (Herrmann, et al., 2022). Bringing the proposed improvements into reality is a task for change management. Possibilities for process improvement can also be identified with process mining (R'bigui & Cho, 2017). AI can also contribute to the analysis and improvement of processes. BPM systematically aims at quality assurance and reliability (Stravinskiene & Serafinas, 2020), for example to eliminate undesired outcome as early as possible before unnecessary work is invested.

Only after the processes have been made as efficient and reliable as possible, projects of supporting the process with technology, especially workflow-management systems, should start (Hammer, 1990). These systems are the technical instantiation of BPM since they automate and help to monitor and control pre-specified business processes. They support running the processes of similar tasks in a standardized way. The standardization of processes is seen as a hindrance to the flexibility required to deal with exceptional cases (Saastamoinen & White, 1995). Projects of introducing workflow-management systems possibly go through adaptations if limitations for standardization become apparent (Herrmann & Hoffmann, 2005). Consequently, workflow-management systems have to include possibilities that help to adapt them.

Business processes are not only relevant for organizational practices of task handling within a company but also for the interaction between networking organizational units. This increases the need for flexibly reacting to changing conditions.

**Relevance for human-centered AI:**

With respect to organizational practice, AI will increasingly not only be applied as tools at individual workstations but be integrated into workflows as components that interact with humans and conventional IT-infrastructure. Thus, the concepts of business process management and the implementation of workflow management systems have to be reconsidered with respect to technology that possibly can take over autonomous decision making and quality assurance. AI can help to analyze processes and to identify organizational ways of improvement in the early stages of projects. AI can also help to support flexibility by self-adaptation or by making recommendations (Herrmann & Nolte, 2024). The potential autonomy of AI for flexible decisions within workflow management systems increases the challenge of exercising oversight by human actors in accordance with HCAI. Oversight of workflows that include AI components needs to be supported by appropriate functions, e.g., through the possibility of intervening into ongoing business processes. Intervening use is a HCAI-related concept in which users only occasionally or temporarily interrupt an automated process in order to take control or adjust parameters (Herrmann, 2025b); interventions or their effects are limited to selected time slots.

## 3. The cases: Predictive Maintenance and AI

The cases described in this Chapter stem from the field of predictive maintenance (PM). Predictive maintenance is considered to be one of the most important AI applications for Industry 4.0 (Garcia et al., 2020). It is used for process technology such as energy supply as well as for infrastructure facilities (Timofeev & Denisov, 2020).

PM is defined as a "technology to recognize the condition of an equipment to identify maintenance requirements to maximize its performance" (Zonta et al., 2020, p. 49).The goal of maintenance is "to keep physical assets in an existing state," while predictive maintenance (PM) is a way "to view data and does not necessarily require a lot of equipment" (Levitt, 2003, p. 91). Maintenance and repair need anticipation of problems and planning. The goal is to avoid unexpected malfunctions as far as possible, to reduce the intervals and duration of downtimes, and to procure or replace spare parts only if necessary and as late and infrequently as possible. The description of case A provides more details with a diagram of how a PM triggered maintenance workflow can look.

PM is usually considered a method that can benefit from the employment of AI, especially machine learning. It has become technically possible to evaluate large amounts of sensor data provided by the systems in real time. Additionally, one can examine datasets produced over long periods of time. Thus, it became possible to detect correlations that were previously undiscovered, and combine them with other data, for example, from the Production Planning System. With the availability of this data, new advancements in AI contribute to the success of PM (Mahale et al., 2025). Provided that sufficiently large and fine-grained datasets are available, ML enables more accurate anomaly detection, anticipating failures before they occur and allowing corrective maintenance to be scheduled in advance (Kroll et al., 2014). In this way, ML-based PM systems can reduce unplanned downtime and manage maintenance cycles more effectively. By focusing on the capabilities of AI, organizational considerations as a success factor tend to slip into the background.

Two cases were analyzed through interviews and group discussions. In case A, the empirical investigation was focused on the project members who were establishing the PM system. Various opinions and viewpoints were elicited that characterize different options for how PM can be organizationally embedded. In case B, an AI-based outlier detection was already established. It became possible to understand how the interplay between service technicians and data analysts evolved and which expectations were discussed with respect to future development.

The empirical investigation of the potentials for AI usage and AI evolution reveals different scenarios and patterns of organizational practices that will be presented in Section 4. Both cases reveal that it is a promising strategy of appropriation to combine deep learning AI systems with rule based systems (Mende et al., 2022) and rule extraction (Barbado et al., 2022).

## 3.1. Case A – Projecting advanced PM

This case analyzes the implementation of PM in the chassis manufacturing division of a German car manufacturer employing approximately 1,500 staff. The division undertook a PM project to enhance maintenance efficiency and reduce unexpected equipment downtime. The study drew on company documentation as well as four in-depth interviews conducted with key personnel: a plant operator, a master craftsman, a systems specialist, and a planner. These interviews were transcribed and analyzed with respect to categories of challenges and opportunities in PM implementation. In total, 77 specific challenges related to the integration of PM were identified (Herrmann, 2020).

At the heart of the project was the construction of hypotheses to determine under what circumstances PM warnings should be triggered. These hypotheses were built from fluctuations and outliers observed in continuous streams of sensor data. If certain thresholds were violated, the PM software issued a warning (see Fig. 1). The plant operator, as the primary recipient of these warnings, was responsible for deciding on their relevance and whether further action was necessary.

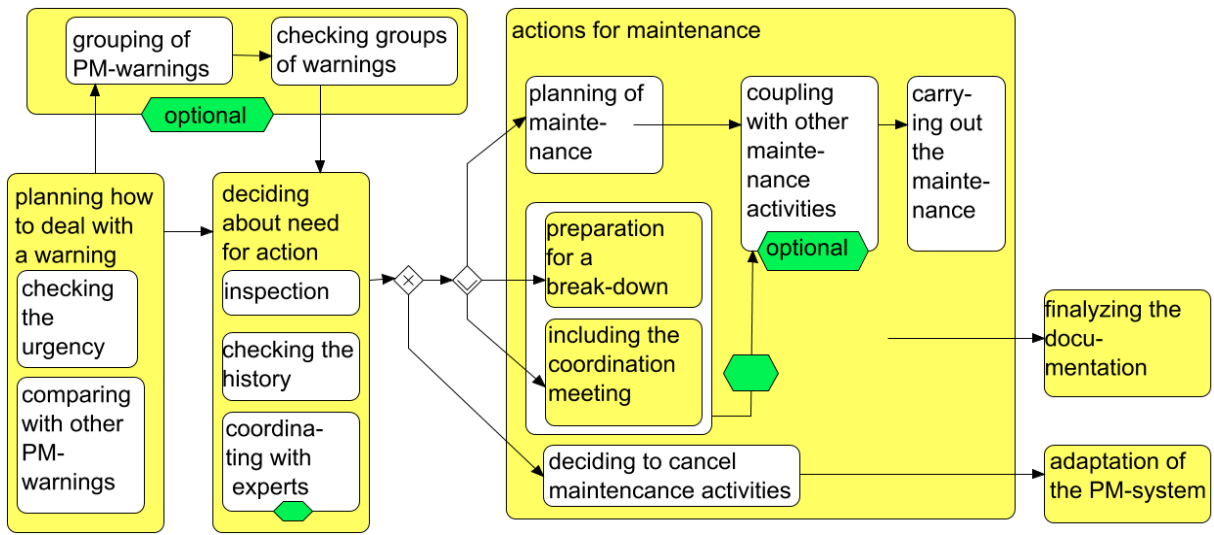

Fig. 2: Workflow of handling a PM-warning (Herrmann, 2020, p. 302)

Initially, the approach relied heavily on staff observations and accumulated expertise to formulate the rules interpreted by the PM software. However, this method soon revealed its limitations. It became clear that a machine learning (ML) approach could enhance the system by replacing the manual process of hypothesis-building.

The testing phase at the beginning of the implementation revealed substantial challenges. Chief among them was the high number of false positives, producing warnings without real cause. They not only burdened the plant operators but also risked eroding trust in the PM system. Each time a warning was issued, the operator had to conduct an on-site inspection, which was time-consuming and resource-intensive. To cope, operators often pooled and prioritized warnings, deciding which were urgent enough to warrant immediate action.

The plant operator's task extended beyond technical checks to coordinating resources, negotiating repair schedules, and consulting with other experts when necessary. If a warning proved unfounded, or if a breakdown occurred without prior notification, adjustments to the underlying hypotheses and thresholds had to be considered. These adaptation processes were often collaborative, involving not just individual operators but also teams of specialists.

Such collaboration was critical in fostering organizational learning, as operators and technical experts jointly assessed the causes of anomalies and determined system refinements.

In practice, plant operators' daily routines already encompassed starting and stopping the plant, preparing production runs, ensuring the supply of parts, and carrying out inspections and basic repairs. With PM, these responsibilities were expanded by a second layer of tasks, as depicted in Figure 1. These included not only responding to warnings but also documenting all actions taken. Documentation served a dual purpose: as a knowledge base for handling future warnings and as input for improving an ML system.

From an organizational perspective, the PM implementation illustrated a bottom-up approach. Rather than being managed solely from the top, adaptations and improvements were rooted in the daily experiences and expertise of the plant operators and their colleagues. This approach ensured that the overall socio-technical system could be continually refined in practice, though it also carried the risk of nontransparent decision-making processes. With regard to this risk, management discussed the option that the adaptation of PM warnings or recommendations should only be conducted by selected key users.

## 3.2. Case B – Monitoring and Analyzing the State of Pump Systems

This case examines the customer service (CS) division of a pump manufacturer, focusing on the integration of machine learning (ML) into service processes (Herrmann, 2023). Pumps are not only delivered but also installed and commissioned by the manufacturer at customer sites. To ensure quality and consistency, the company has established systematic process management. One central process governs the sequence from the request for an offer through to the final installation of the pump system. This formalized workflow, supported by software-based form filling, ensures that all necessary information is complete and consistent before the pump is put into operation. This structured information exchange contrasts with the more agile, project-oriented communication used for innovative initiatives, such as the use of AI and its continuous refinement.

Customer service processes begin once the pumps' sensors are operational and capable of transmitting data from the customer's site. At the outset, the CS team reviews the first data streams to gain an understanding of the customer's equipment. A meeting with the customer and its plant operators follows, where the features of the monitoring and analysis (M&A) system are explained in the context of the newly installed equipment. A customer may choose to have the plant operators use the M&A tool themselves; however, many clients prefer to rely on the manufacturer's CS team for professional support.

The M&A tool reflects the real-time operation of pumps and provides several core features:

- **Real-time data delivery** on operational conditions at the customer's site.
- **Statistical analysis and graphical visualization** of aggregated data to support interpretation.
- **Rule-based notifications** triggered when thresholds are exceeded, signaling malfunctions or potential breakdowns. These thresholds are derived from prior experience with malfunctions.
- **Machine learning-based anomaly detection**, employing unsupervised methods to identify outliers. This allows the detection of problems not captured by rule-based systems, especially those reflecting previously unknown phenomena. Such detection is critical when deviations might escalate into cascades causing major breakdowns.

To make effective use of M&A, the CS team integrates two domains of expertise: technical specialists such as electricians and mechatronic engineers, and data analysts. Their collaboration ensures that technical anomalies identified by the system are properly understood and translated into actionable steps for the customer. Importantly, the M&A system itself is continuously improved. The CS team therefore works closely with the development team, with data analysts forming the bridge between the two groups (see Fig. 2). Regular weekly meetings support the iterative development and refinement of M&A features.

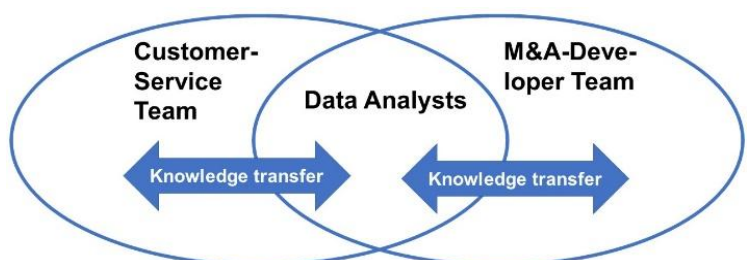

Fig. 2: Data analysts at the intersection (Herrmann, 2023, p. 252)

The CS team becomes active through several types of triggers:

1. **Customer-reported problems**, which may or may not be confirmed by data.
2. Indicators delivered by the M&A component,
    - plotted data trends that suggest deviations from normal patterns.
    - **Rule-based warnings** that are issued when defined thresholds are violated.
    - **ML-based anomaly detection**, which reveals irregularities that would otherwise be invisible.

In each of these cases, data analysis signals the possibility of malfunctioning. Human decision-makers must then evaluate the situation and decide on necessary actions, often consulting the customer. These decisions may also prompt adjustments or improvements to the M&A component itself, ensuring that the system evolves with the operational challenges encountered.

By embedding ML-supported anomaly detection into the service process, the pump manufacturer not only strengthens its preventive maintenance capabilities but also enhances customer trust. The collaboration between technical and analytical experts, anchored in systematic communication structures, ensures that the AI-driven system supports both reliable pump operation and continuous organizational learning.

## 3.3. Keeping the Organization in the Loop

Based on case A, Herrmann and Pfeiffer (2023a) developed the concept of keeping the organization in the loop. Case B helped to substantiate this concept. Through empirical analysis, they identified ten types of interacting organizational practices that need to accompany human-centered AI which they understand as an approach to keeping the human in the loop. Figure 3 represents the eight most essential types of organizational practices by describing their purposes:

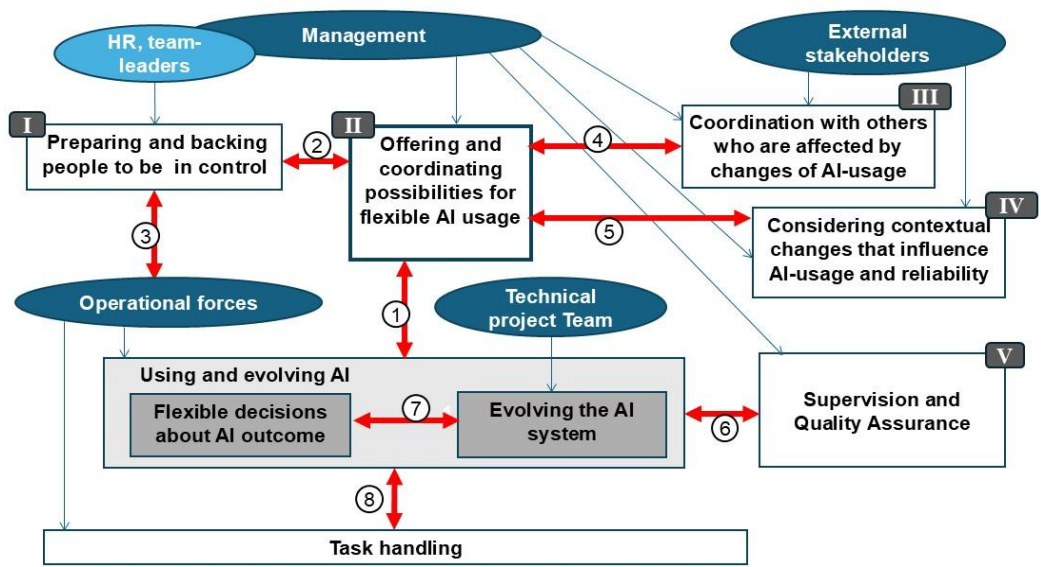

Fig. 3 Organizational practices and embedding of AI

1. In the center of Fig. 1 is the managerial task of offering and coordinating the possibilities for flexible AI-usage. This managerial coordination provides rules that guide the operational forces by clarifying what kinds of flexibly adapting AI outcomes can be achieved by whom and under which conditions. The rules for coordinating the handling of AI need to be transferred into organizational practice (Section 2.2). It combines AI support with the handling of the original tasks and must ensure that human decisions are still involved. This requires, for example, allowing sufficient time to consider explanations for an AI outcome and to check its quality. Furthermore, management has to coordinate the interplay between using AI and its evolution.
2. Making autonomous usage happen not only requires that people be allowed to flexibly decide on using AI outcomes. They also need to be prepared and encouraged to do so. Thus, the coordinative measures have to be coupled with HR-development and be supported by team leaders so that flexible handling of AI results takes place, such as rejecting or adjusting them. It must also be part of organizational practice that both AI's and employees' capabilities are continuously developed and that everyone is encouraged to contribute to this learning process.

3. Preparing, encouraging, and backing the operational forces to autonomously use AI requires intensive exchange with them. Adaptations of or reflections on AI outcomes, as well as shortcomings of AI, can be documented to provide a basis for further reflection. This reflection should not only take place individually but also collaboratively within teams and with supervisors. Employees need to know that their peers and supervisors will appreciate it when they counteract or challenge the performance of the AI.
4. Since organizational units communicate with other stakeholders, these stakeholders also have to be informed about the possible adaptation of AI results that might affect them. It must be specified who in the environment will be influenced by dealing flexibly with AI and whose interests might be affected, etc. For example, if a PM-system triggers an alarm and the plant operator overrides this alarm, then a service team must be able to understand this decision.
5. The environment of an AI application is subject to continuous changes that need to be considered by management when coordinating the possibilities for usage and evolution. Changes may be triggered by the market or by new technologies. Changes in contracts or laws might require standard procedures that prevent certain types of decision making that violate regulations. However, flexible adaptation might even be necessary to allow for a quick reaction to new legal requirements or ethical discourses about AI. Besides shortcomings of a certain AI application, the continuous change of the context of AI usage is a further reason for organizational practices that enable flexible decisions and the continuous evolution of AI.
6. From an organizational point of view, using AI, its evolution, as well as the interplay between both, requires organizational practice of close supervision. This includes testing AI and its reconfiguration as well as evaluating whether the adaptation of AI outcomes has been reasonable or not. In particular, organizational practice needs to consider whether the same kinds of adaptations happen too often and should be avoided through reconfiguration.
7. Flexible handling of AI outcomes and the evolution of AI are intertwined through organizational practice. Adaptation of an outcome influences how an AI system will evolve or be reconfigured, and the reconfiguration itself changes the nature of the outcome and the possibilities for understanding it and dealing with it. These changes must be considered by the organizational practices that surround AI usage, for example, to allow for experimenting with preliminary adjustments before implementing them permanently.
8. Using and influencing AI need to be closely related to the task handling that is to be supported by AI. The task handling procedures might have to be adapted and adjusted to AI usage. For example, employees can be encouraged to first consider for themselves what decision they would make before looking at the AI's proposed decision (Fogliato et al., 2022). Such a measure makes reflection on AI results and their possible rejection more likely and contributes to the calibration of trust in AI.

The description of these eight organizational practices clarifies how keeping the organization in the loop looks like and that it is constantly relevant – not only at the beginning when AI is implemented. Herrmann and Pfeiffer investigated the extent to which the findings they had made in the field of predictive maintenance could be generalized. To this end, they analyzed literature on the use of AI for medical image detection (Herrmann & Pfeiffer, 2023b). Their results confirm that the management activities shown in Fig. 3 are also relevant in this area. Their analysis leads to a more detailed description of these management activities (Herrmann & Pfeiffer, 2023b):

I    Coordination of AI-related tasks
  a. Defining workflows and new tasks of using AI within workflows of original tasks
  b. Defining the ecology of relevant roles
  c. Reflection of interdependencies between contextual factors and AI
  d. Checking for integration of all supports
  e. Adding resources
  f. Preparing the interplay between quality assessment and adaptation of AI
  g. Supporting the exchange of stakeholders' perspectives

II    Leadership and HR as enablers
  a. Aligning roles, people, and tasks
  b. Employing and developing human experience and capabilities
  c. Preparing users to contribute to adaptations
  d. Preparing staff for the shortcomings of AI and for supervisory tasks
  e. Preparing staff for dealing with the subjective attitudes and experience of stakeholders, particularly customers

III  Coordinating with the external world
   a. Regulating access from the outside
   b. Identifying and observing various and changing external influences to be addressed
   c. Regulating the sharing of risks and benefits, e.g., exchanging data and technology
   d. Building strategic partnerships
   e. Supporting and regulating cross-institutional exchanges of underlying data

IV  Dealing with contextual factors and changes
   a. Assigning measures and resources to various types of prediction
   b. Tracing the causes of problems
   c. Deciding about risk-benefit trade-offs, e.g., for exchanging data and technology
   d. Organizing multilevel long-term changes
   e. Observing and predicting societal changes

V  Supervision and quality assurance
   a. Organizing supervision routines
   b. Organizing tests when new algorithms are deployed
   c. Regulating the modification of ML algorithms
   d. Ensuring quality for data to be used and shared
   e. Organizing the involvement of external expertise.

During the interviews and group discussions in cases A and B, it became apparent that the implementation of these management activities is not a matter of course. Management was considering various options for how it could engage in the implementation of AI. Some hoped that it would be sufficient to implement AI as a new tool with minimal organizational adjustments. Others believed that only a few organizational changes were necessary at the beginning. Still others recognized that more communication and quality assurance were needed in order to exploit the potential for innovation. In the following section, we outline the various options that were discussed by describing ten patterns for how AI could be implemented and used.

# 4. Socio-Technical Patterns of Organizational Practices Derived from the Cases

Section 3.3 describes organizational patterns with respect to their purposes and contribution to keeping the organization in the loop. However, it remains unclear what these patterns might concretely look like. Therefore, in what follows, ten patterns are described derived from the options that were discussed during the empirical investigation. A variety of different ways to organize the implementation of AI, its usage, and the continuous development of its socio-technical integration can be identified. The most significant variations are explicated in the following patterns. They present different organizational solutions for the socio-technical integration of AI. The differences are decisive for the degree of human-centeredness that can be realized with respect to assigning relevant tasks to humans, supporting their development, and autonomy, etc. (see Section 2.1). The various options represented by the patterns do not support HCAI to the same extent. However, it can be indicated how different organizational options lead to different levels of compliance with the HCAI goals and principles described above.

## 4.1. Organizational Embedding of AI

### 4.1.1. AI as a Component of Control-Oriented Workflows

An AI-based PM is a component of an extensive workflow that starts with the sensing of outliers in the state-transitions of a plant and ends with several possible states such as the completion of repair measures, preparing actions of repair, adjusting the PM system, etc. Depending on the assumptions about AI, there are two main options for how humans can be integrated into such a workflow that includes AI components. (Section 2.7).

Option A): Some interviewees supposed that nearly every case is a good case where all of the warnings or outlier detections refer to real problems. Subsequently, AI could propose a valid solution for how to overcome the problem and several succeeding steps can be initiated: depending on the predicted urgency, a time slot for repair measures is identified, staff is assigned to the repair activities, spare parts are ordered form storage and/or from a supplier, and the success of the repair work is finally checked. All these steps have to be accompanied by a human agent for the purpose of quality assurance.

With regard to HCAI, it can be noted that this remaining task is limited to that of a supervisor who approves the transition from one step to the next. He or she must monitor a large number of AI-based PM alerts for a series of various equipment. This can lead to monotonous work, where routine-based overtrust can prevail (Section 2.2). The interviewed plant operators have assumed that such a higher-level supervisor, who is responsible for an entire series of robot-operated workstations, does not have the experience gained from the details of daily plant maintenance. Thus, the HCAI goal of keeping those people—in this case, the plant operators—in the loop who have the relevant knowledge and experience will not be achieved. Possibly, as a contribution to HCAI, the activities on the human side can be enriched, e.g., by the task of negotiating with repair workers to find someone to show up for a weekend shift.

Option B): In Case B (Section 3.2), the data analysts did not suppose that all detected outliers represent serious problems that one has to react to. Since false positives are frequently possible, the outlier detection is interwoven with a more complex workflow that includes different contributions from the human side. Triggering the workflow is based on two streams of observation: one through technical monitoring via sensors and one through continuous observation of the plant by the operator. If either of them suspects a problem, the validity of this problem can be checked by interacting with the AI system and by performing an additional inspection at the plant. Depending on how certain one is that there is actually a problem, repair can be initiated, only prepared, or rejected as unnecessary. In the case of a real problem, a solution may be proposed by either the AI or the human operator or both of them. Then it needs to be checked how valid or efficient a solution is before the steps for repair are initiated. The workflow must be further extended if a team of specialists needs to be consulted. Furthermore, for each observed shortcoming of an AI-based proposal, it has to be decided—by an additional workflow—whether the underlying AI system or workflow needs to be adapted or not (Section 2.3).

With respect to HCAI, option B) offers a higher potential: Involved people have more autonomy. They know that they can contribute to the improvement of AI-based findings based on their knowledge and experience gained on the job. They can feel like agents who drive the improvement of the socio-technical integration of AI. Thus, they contribute to continuous organizational learning. Considering the principle of human led collaboration (Xu & Gao, 2025), it turns out that this is not only relevant for teamwork but also for workflow-driven collaboration. In Option A, AI is the driver, and the human side has only the possibility for occasional intervention, while option B offers the opportunity for humans to drive the progress of the workflow.

### 4.1.2. Top-down Regulation Aiming at Centralized Control

The middle management of Case A supposed that PM produces warnings and maintenance requests that may frequently be inaccurate or economically questionable. Therefore, they began considering organizational concepts for how the AI outcomes could be adjusted to reduce these deficits. They also recognized the potential need for PM systems to be continuously customized. For example, the adjustment can be done for single cases of a warning or with a sustaining effect, e.g., by formulating rules or storing documents for retrieval-augmented generation (Cheng et al., 2025) that reduce the number of false alarms. An organizational practice was proposed to regulate this adjustment or customization in such a way that only a few employees, so-called key users, are authorized to make adaptations. They would also be the only ones responsible for deciding to whom maintenance tasks triggered by PM are assigned and when. This is intended to counteract the risk of employees manipulating the PM software so that they themselves only have to carry out a few maintenance tasks. For example, it was assumed that such manipulation could happen by reducing the thresholds to be violated before a warning is issued.

This management pattern of top down regulation represents a type of centralized control requiring that all measures leading to the postponement or delegation of maintenance tasks be authorized by the key users. Therefore, the steps of adjusting AI outcomes or customizing AI are precisely prescribed. This opens up the possibility of centrally monitoring the workflows related to AI-based PM usage. Thus, the design of the workflow for handling PM outcomes or customizing them becomes a subject of centralized quality assurance (Section 2.7).

With regard to HCAI, it can be said that this pattern leaves important tasks to humans and fosters being in control, but only for a few—namely, the key users. In the interviews, concerns were expressed that the key users could be overburdened, as they lack detailed knowledge of the equipment and their decisions may be guided by mistrust rather than expertise. Thus, the top-down pattern would not allow for human oversight and human-led use of AI to the same extent as teamwork including all available experience.

### 4.1.3. Bottom-up Improvement

In particular, plant operators raised concerns that key users or key decision-makers might be too far removed from the specific characteristics of the plants that are the subject of the PM. In contrast to the previous pattern, the task of continuously adapting or improving AI can therefore be organized in such a way that it builds on the experience and work processes of the employees directly involved. These are the staff members engaged in the monitoring, maintenance, and repair of a system, as well as in tasks such as setup, quality control, and similar activities. They can propose changes to the AI directly and implement them immediately—at least on an experimental, preliminary basis.

Under these conditions, a decision-making process needs to be organized to determine whether and how such preliminary changes can possibly become permanent adaptations. This decision-making process is conceived less as a top-down management task with key users and key decision-makers and more as a team responsibility, involving experienced machine operators, their immediate supervisors (foremen), and specialized workshops (Section 2.5). The advantages and disadvantages of potential adjustments are discussed within the team. Where appropriate, past cases are reviewed to verify whether the AI adjustment is truly beneficial for a multitude of cases. Experimental phases can also be introduced to assess the success of an AI modification.

This approach is more in compliance with HCAI than top down regulation. It delegates (Section 2.4) quality assurance of AI to those who are directly involved in the monitoring and maintenance of plants and seeks to make the broadest possible use of their existing experiential knowledge (Section 2.1). With this pattern, those employees are leading the interaction with AI who are directly concerned with and experienced in the problems that occur. The bottom-up approach helps them directly maintain their interests while employing AI. However, it implies a high degree of dynamism and carries the risk that nontransparent, informal decision-making processes may occur—processes that could be perceived by management as a loss of control.

### 4.1.4. Appropriation: Combining AI-based Analyses with Conventional IT

While Case A focuses primarily on the PM-system, Case B concentrates from the very beginning on intertwining AI with conventional IT infrastructure. The AI was intended to supplement other tools, such as data visualization or rule-based software for detecting deviations from standard operating procedures. Thus, AI was primarily introduced to identify previously unknown deviations, that the rule-based software could not detect from normal operations. Accordingly, the customer service team analyzes the underlying factors that cause AI to report an outlier. For this analysis, service technicians and data analysts work together. If they understand the reasons that triggered the detection of a deviation, their organizational practice results in two succeeding steps:

a) they take further steps to secure the customer's pump system against malfunctions.
b) they incorporate the AI's detection of an outlier by extending the rule-based software with additional rules.

The second step can be considered a means of evolving the technical infrastructure based on the AI outcome—not necessarily the AI itself, but the rule-based system. Using the rule-based software to predict problems is easier to understand, since the rules that triggered an alarm based on incoming sensor data can be displayed when needed. This transformation of AI-based reasoning into rule-based reasoning might be considered unproductive extra work, but for the staff, it is a method of employing the AI system with the effect that the service team and data analysts feel in control. This can be considered an appropriation (see Section 2.1) of the AI system. Similarly, Case A exemplifies that the use of rule-based systems opens the possibility of experimentally modifying the thresholds used in the rules. Apparently, the usage of AI to solve PM-problems and the continuous improvement of AI and its related infrastructure are inseparably interwoven.

This pattern strongly supports the goals of HCAI. Humans are largely in control and can systematically seek to understand the reasons behind AI outcomes. They use AI to further develop their understanding of data analysis and about how pump technology work at the customer's site. The process of appropriation is a matter of teamwork, where data analysts and service technicians work together. Thus, the pattern represents an organizational practice of socio-technical integration to bring HCAI into reality.

### 4.1.5. Explainability and Accountability as an Organizational Task

The previous sections (4.1.3, 4.1.4) provide examples of how users in Cases A and B seek to understand, use, and influence AI outcomes. This endeavor is closely linked to the possibilities offered by explainability (XAI), which is a basic feature of human centered AI (Ehsan & Riedl, 2020). Most concepts of XAI frame it as an interactive feature between the user and AI. The common discourse assumes that AI software should be able to explain to the end user the background of the results it generates. Thus, users can understand the decisions AI suggests, so that they can assess the quality of the AI's performance and build trust in its outcomes.

Supporting the use of explainability requires a pattern of organizational practices in which employees collaborate to combine different areas of expertise as in Section 4.1.4. Explaining AI turns out to be teamwork that requires communicative effort (see section 2.2 and 2.5). It is a collaborative endeavor—a kind of organizational practice in which several people function as a team to integrate relevant expertise from different perspectives. In Case B, data analysts help service technicians understand why an outlier was detected and which data constellation was decisive for its identification. The service technicians, and possibly engineers responsible for pump development, can help determine whether the identified outlier indicates a serious problem or a normal, though rare, phenomenon.

Explainability at all—whether it is a task of individual users or a collaborative endeavor—requires that organizational practices allow sufficient time for considering explanations. This time has to be made available during regular task handling. Management must support (Section 3.3) the development of conventions that clarify when and how much time, seeking an explanation for an AI outcome is possible (Section 2.2). An important aspect is that organizational practice encourages reflection on the contribution of AI to the execution of a specific task. Communicative routines are integrated into this pattern to use explanations to support accountability and organizational learning (Section 2.3). With this pattern, explanations of AI behavior serve not only the immediate purpose of verifying or quality-checking an AI result. They also reveal the underlying contexts that can be used to improve technical systems (Section 2.3) or that can serve as a basis for explaining certain phenomena to customers in a way that they can understand and know how to respond to. This, in turn, implies an organizational practice in which a human intermediary is involved in conveying explanations to those who are affected.

Apparently, this pattern outlines explainability as a socio-technical phenomenon that directly contributes to HCAI. It is the basis for complying with the principle of transparency. It is not limited to achieving an explanation but also includes the further use of it: reflecting on the explanations helps meet the requirement for accountability, and conveying the explanation appropriately to those who are affected may help recognize and avoid biases as a contribution to fairness.

### 4.1.6. AI Usage as a Driver for Organizational Learning

The described transfer of AI results into the rule-based software (Section 4.1.4) is a kind of appropriation that can build a basis for organizational practices of organizational learning (Section 2.3). Such a pattern of organizational practice realizes the elaboration of new rules as a special way of documenting and disseminating new knowledge. This possibility was recognized and discussed in Case B. Analyzing the background of an AI-based outlier detection was initially used as a method to improve collaboration between data analysts and service technicians in the detection of problems and the maintenance of pump systems. This organizational practice supports learning processes: the service technicians gain an increasingly better understanding of how to approach data analysis, while the data analysts acquire deeper knowledge of the technical specifics of the pump systems.

The scope of the organizational practice can be extended to trigger learning processes that go beyond the team level, with the goal of informing and improving the work of other departments in the company. While the learning in the team was grounded in an organizational culture of informal organizational practices and communication (Section 2.2), the dissemination of information to other departments was regulated by a formal reporting system. An effective pattern for company-wide organizational learning has to overcome or at least complement formal reporting to make it easier to share information instead of adhering strictly to certain documentation standards. From the viewpoint of organizational learning, the insights of the service team should freely flow into the departments responsible for the design and manufacturing of pump systems. Thus, the newly detected rules can help improve the construction of pumps—for example, in terms of durability, fault prevention, or the efficient execution of maintenance work. Consequently, the organizational practice of handling outlier detections should not just locate and eliminate a potential source of malfunction but also contribute to understanding the underlying causes. Achieving this understanding needs to be appreciated as a task involving collaboration across several departments and with external stakeholders (Section 2.6). From this perspective, the use of AI is not limited to the narrow focus of interaction between the direct end user and the AI system but instead extends its influence to broader organizational levels.

The promotion of organizational learning is not directly formulated as a goal by the HCAI community. However, it contributes indirectly to HCAI. It keeps people in the loop, as employees contribute to organizational learning through their individual learning progress. The example of organizational learning demonstrates how HCAI can contribute to innovation processes and bring long-term benefits.

### 4.1.7. Embedding AI-based Task Handling into Communicational Exchanges

The organizational embedding of AI-based decision making, explainability, organizational learning, knowledge management, and the continuous adaptation of AI each include communication processes between various stakeholders. Case A reveals that these processes can become rather complex and require a kind of communication-oriented organizational practice.

Predictive maintenance (PM) does not always produce unambiguous forecasts of potential malfunctions that can be clarified by one individual alone. This requires a socio-technical solution and an organizational practice that involves various specialists. Meetings must be organized to clarify how relevant the warnings are, whether it is justified to initiate countermeasures, and who should be involved in the repair work and when. One might decide, for instance, that a part of a machine assessed as critical should not be replaced immediately, but rather during scheduled repair work carried out over the weekend. In such a case, the required replacement part would be procured in advance and stored in close proximity to the system. Such a complex, collaborative decision has to be supported by the organizational practice of this communication-oriented pattern.

It is evident that multiple contributing actors need to be identified and involved. Information must be made available across shifts, and the background of the warning and the potentially defective part must be documented and kept accessible. For the purposes of organizational learning, the entire process must be traceable (Section 2.3). From this perspective, an organizational culture needs to be developed (Section 2.2) in which the use of AI for predictive maintenance is integrated into communicative workflows and meetings. In addition, a kind of technical infrastructure has to be employed that supports formal as well as lightweight kinds of coordination and documentation. More advanced socio-technical practices can include AI to support communicational processes, e.g., by selecting who should be involved, scheduling meetings, or coordinating a decision-making process about whether an expensive part should be ordered.

With respect to HCAI, it must be emphasized that it is a basic human need to communicate with others during work (Section 2.1). Organizational practice must ensure that this communication is possible. Furthermore, exercising control and autonomy are the result of social interaction in which people confirm to each other that such behavior is possible and acceptable and inform each other about how to achieve it.

## 4.2. AI in Different Roles

While the organizational practices described in Section 4.1 were directly inspired by the empirical work in Cases A and B, the various roles of AI referred to in this subsection are mainly discussed in the literature. However, roles such as teammate (Zhang et al., 2021), powerful tool (Shneiderman, 2022), or sparring partner (Herrmann, 2025a) can be easily connected to the case studies and enrich the scope of possible organizational practices. Organizational practices that include these roles can contribute to the socio-technical design of HCAI. They need to be considered in the discussion (Section 5) of the barriers and enablers that influence the implementation of HCAI.

### 4.2.1. AI as a Teammate

One proposal for embedding AI into a socio-technical solution is to establish it from the outset as a team member (Zhang et al., 2021)—one that is integrated into the division of labor and is also involved in relevant communication and decision-making processes. These could also include coordinating tasks and supporting continuous improvement. In Case A, the underlying organizational workflow was drafted with the following projection: the plant operator receives a warning, checks the plant, and decides whether the situation is clear or further clarification is needed via communication in a team meeting (Section 4.1.7). For such a team meeting, the plant operators simply introduce the output of the PM-system. Possibly, they can interact with the PM-system during the meeting. From this perspective, one can imagine an organizational practice where an AI-based PM system is directly included as a teammate (Section 2.5): it would join the meeting and justify why it has issued a warning, propose solutions and contribute to the discussion about the available options for handling the warning.

With respect to Case B, AI could itself make proposals about how to adapt the rule-based software. Such a scenario becomes more plausible with LLM-based GenAI systems that can enter natural language conversations. The practice of exercising oversight over AI would be converted into critical discourse with the new AI team member. Integrating such a new team member into an organizational culture, in which plant operators and technicians are used to a habit of controlling and mastering technical artifacts will require thoughtful attention to prepare and accustom the staff to the new teammate (Section 2.2). Organizational practices will include frequent reflection and negotiation on how well the new team member performs and which tasks it could take over (Section 2.3). The negotiation will address the question of which competences one will assign to the AI teammate (Section 2.1), e.g., whether it will be allowed to take over the coordination of the repair date and the assignment of the repair team.

Such an extension of competencies would reduce managerial work on the one hand but—with respect to HCAI—increase risks of losing control on the other. The advantage for humans could be that interaction with AI directly becomes a matter of teamwork and that end users do not have to pass on their recent experiences with AI to others at additional expense. Full integration as a team member would imply that the AI system could take part by itself in the reflection and negotiation on the distribution of tasks. To comply with HCAI, the relevant organizational practices must clearly distinguish between AI's ability to participate in communication and its status, which is not of a social nature. AI can participate in discussions, but the human voice is crucial for the final decision, for which humans need to be responsible and in the lead (Xu & Gao, 2024).

### 4.2.2. AI as a Powerful Tool

Shneiderman (2022) portrays AI more as a powerful tool rather than an intelligent agent, teammate, or co-active collaborator. Such a tool can augment human cognitive processes and creativity. With respect to the example of predictive maintenance, one can imagine organizational practices where the PM system is used simply as a tool with which the plant operator autonomously (Section 2.1) draws her or his own conclusions about the state of the plant. There would be no direct need for discussions with other teammates or experts about how far the issued warnings might be appropriate, what kind of explanations or rules might be hidden in the AI-based reasoning, or whether the sensor data is sufficient and correct, etc.

The plant operator develops her or his assessment of the situation, and no further discussion or coordination is planned for the organizational practice. Only the phase of implementing an AI-based PM would require some new organizational practices, such as providing AI-related training (Section 3.3). Such use as a powerful tool would match quite well with the culture of technicians. Additional time to deal with explainability, explore the potential and limits of AI, contribute to the continuous evolution of AI would not need to be included in the organizational practice of using a powerful tool. The organizational practice would be based on the assumption that the tool provides only functions that are reliable, without routines intended to compensate for its imperfection. All other necessary functions are contributed by human intellect. It could be interesting—but not necessary—to document cases of using such a powerful tool to contribute to knowledge management and organizational learning (Section 2.3).

Such a pattern would keep the organizational effort low. With respect to HCAI, the usage of AI as a tool emphasizes the autonomy of the user. Tool usage is designed to keep users in control after they are trained for it. An organizational practice that does not include routines for seeking explanations or communicating with others might overwhelm the user with the challenge of dealing with the complexity of AI. Thus, compliance with HCAI goals only increases if the pattern of using AI as a powerful tool is combined with other patterns that emphasize communication (Section 4.1.7) and explanation (Section 4.1.5).

### 4.2.3. AI as a Sparring Partner

Case B shows an attitude among the involved technicians and data analysts that assumes AI can make mistakes and that, if this is not the case, they can figure out the reasoning behind the detection of outliers. This attitude implies that the user may be able to—or can learn to be—better than the AI, or at least significantly supplement its results. This attitude promotes a pattern of organizational practice in which AI is employed as a sparring partner (Herrmann, 2025a). This organizational practice envisions AI primarily as a tool for training, for improving a team's performance or for learning on-the-job rather than for immediate task execution. The use of AI as a sparring partner can be organized in those fields where employees or management believe that staff in the organization's core business are possibly capable of competing with AI—or at least can learn how to surpass it. Drawing on the analogy of sports training, the organizational practice is based on the assumption that a sparring partner is a person with strengths similar to those of the employee being supported. AI as a sparring partner is a friendly opponent used for competitive interaction in which someone attempts to solve problems better than the opponent to improve her or his skills.

Also, in Case A, the technicians feel challenged by the warnings provided by AI and learn to deal with them critically. An organizational practice that uses AI as a sparring partner must always provide an opportunity for users to solve the problem at hand based on their own expertise. A typical workflow would be that PM gives a warning, and subsequently the user thinks about how to manage the problems behind the warning. Only afterwards is the AI system asked to provide a proposed solution. Then the plant operators can compare their idea with this proposal and find out which one is more promising or how both approaches can be combined.

Organizational practice that pursues this concept of competence development does not use AI primarily to make knowledge content retrievable and deliverable. Instead, it focuses directly on employees' problem-solving abilities and their belief in their own self-efficacy (Section 2.3). Furthermore, by comparing their own solutions with those of AI, employees can not only evaluate their own expertise but also that of AI. This is implicitly an approach toward trust calibration (Okamura & Yamada, 2020) that helps avoid overtrust or undertrust (Section 2.2).

Considering HCAI, using AI as a sparring partner acknowledges the possible strengths of human users and helps them assess or improve their capabilities to stay in a leading role when interacting with AI. The sparring partner approach conveys that AI is not unquestionably recognized as being better than humans.

# 5. Discussion: Dimensions of Varying Degrees of Integrating AI with Organizational Practices

Herrmann and Pfeiffer (2023a) argue for the need to keep the organization in the loop. This need focuses on achieving the goals of HCAI—ensuring quality and avoiding risks. In this context, AI is considered a technology that can hardly be delivered at a perfect level but is mostly a subject of continuous improvement with respect to reliability and efficiency, transparency as well as compliance with ethical guidelines and legal regulations. To support HCAI, quality management and continuous evolution require certain organizational practices that include additional effort. To understand the barriers and enablers of implementing the concept of keeping the organization in the loop, one must comprehend the balance between the effort, risks, or costs of integrating AI with appropriate organizational practices on one side and the possible short-term and long-term benefits for productivity, innovation and well-being on the other. This section explores the chances of making the concept of keeping the organization in the loop a reality by discussing this balance with respect to eight different dimensions. These dimensions consider different levels of effort and completion in integrating AI with organizational practices. For example, integrating AI-components into complex business processes might reduce coordinative work: If spare parts (see Section 4.1.1) are automatically ordered from the supplier and repair work is scheduled without including several levels of hierarchy, organizational effort can be reduced. However, introducing new organizational practices to assess the quality of these processes and to allow for human oversight and possible interventions requires additional coordination. This kind of comparison can be considered in detail within the eight dimensions represented in Table 1.

Table 1: Characteristics of Higher and Lower Organizational Integration of AI

| Organizational effort for … | 1 – Zero or Low Effort | 2 – Medium Effort | 3 – High Effort | 4 – Very High Effort |
|---|---|---|---|---|
| 1. **Integrating AI into processes of task execution** | AI is only used to support separate, individual tasks or procedures | AI is used for collaborative tasks, or an entire job with several sub-tasks is delegated to AI | AI takes over tasks that are integrated into organization-wide business processes | Network-wide business processes are redesigned for optimized integration of AI |
| 2. **Integration into technical infrastructure** | Not planned – AI is only used for individual workstations, e.g., as a powerful tool | AI as an extension or enhancement of other tools | Integration with workflow management systems; further development of other tools | Regular further development of the infrastructure based on learning through the use of AI |
| 3. **AI integration into team structures** | Not intended | AI as a teammate in a few teams | Company-wide integration into teams | AI also takes on organizational tasks in coordinating teamwork |
| 4. **Quality Management (QM) of and with AI** | Considered hardly relevant since AI is regarded sufficiently reliable after the initial implementation and testing. | Regularly necessary at the individual workstation | AI is used for collaborative work; quality assurance as collaborative task | QM continuously takes place as integral part of all relevant business processes |
| 5. **Explanations** | Considered hardly necessary after the introduction of AI | Repeatedly relevant for individual professionals who use AI in exceptional cases | Collaborative explanation in teams or for customers; involvement of specialists and collaborative reflection | Explanations are used organization-wide to justify decisions also for partners |
| 6. **Continuous improvement and further development of AI** | Not a task of the company; the supplier provides new versions | Improvement requirements derived from use and forwarded to internal specialists | Adjustments carried out internally, limited to professionals and small teams | Continuous AI optimization as a company-wide task |
| 7. **Integrating AI usage with reflection in communicational processes** | Not intended | Sporadic, in a few individual teams or exceptional cases | Sporadic, but company-wide, or systematic but only for selected types of tasks | Systematic, company-wide and with partners |
| 8. **Integrating continuous learning with AI** | Considered unnecessary after the introduction phase | Regularly only for individual professionals at selected opportunities | Continuous learning on the job that is integrated into collaborative problem solving of teams | Company-wide organizational learning with AI, including exchange with partners |

The eight different dimensions are derived from the background in Section 2 and the patterns in Section 4. The dimensions are relevant because the patterns in Section 4 do not systematically analyze the potential varieties and implications that influence the success of keeping the organization in the loop and its contribution to HCAI. The eight dimensions are represented in the rows of Table 1. The individual dimensions are explained by progressing from low to very high effort as represented by the four columns in Table 1:

1. **How AI is integrated into processes of task execution**
   Using AI as a powerful tool as described in Section 4.2.2 enables the option of a low level of organizational effort for coordinating task execution: AI is used by a single user for single tasks that do not interact with other tasks. The user mainly needs to be trained in the use of AI and ensuring the quality of single task execution. However, if other roles are collaboratively involved, e.g., for quality assurance, or if there is back-and-forth between the AI-supported tasks and other ones, the coordination effort increases (level 2). This is the case if more complex bundles of subtasks are delegated (Section 2.4) to AI as a teammate (Section 4.2.1) or if AI is integrated into a business process (Section 4.1.1). Then, the need for oversight and interventions also require more deliberate coordination. At the next level of effort, AI is integrated as subcomponent in collaborative workflows, including organizational redesign and continuous oversight and awareness of possible improvements (Section 2.7). An even higher effort is required (level 4) if business processes are redesigned to cover network relationships (see Section 2.6) and aim at coordinated improvement, e.g., in the relationship between customer and supplier.

The effects on HCAI along this dimension point in different directions. At a low level of integration, it is easier to provide transparency and to exercise oversight and autonomy. The more AI permeates all business processes, the more human expertise is required. New sub-tasks arise to monitor the quality of AI-based results and to intervene if necessary. The contributions of at least some people, possibly in the role of key users (Section 4.1.1), become more important with a higher level of effort.

2. **Degree of integrating AI with technical infrastructure**

   As Case B (Section 3.2) and the appropriation pattern (Section 4.1.4) demonstrate, the integration of AI into existing technical infrastructure can have different levels of complexity. Preparing, maintaining and continuously improving technical infrastructure, including AI, require additional organizational practices. If AI is not connected with other technologies, the effort for integration is lowest. As soon as one starts to compare AI results with the outcomes of other tools, the manner and timing of these comparisons, as well as how to deal with their results, must be organized (level 2). More activities (level 3) must be coordinated within organizational practices if tools are adapted based on AI as is the case with the continuous improvement of the rule-based system (Section 4.1.4). A similar effect can be expected if not only business processes but also the underlying workflow management systems are adapted (Section 4.1.1). At a fourth level of effort, the organization learns to improve its technology—such as production methods (see Section 4.1.6) —by exploiting the experience gained through AI usage.

   In this context, HCAI goals are especially achieved for the role of technicians: The more the integration with technology is pursued, the more the role and expertise of technicians are valued. They have a whole set of technical means at their disposal, now extended with AI, that are subject to improvement or that support it. Technicians are then enabled to exercise human-led AI interaction and human-led decision-making with AI and for the improved integration of AI.

3. **Integrating AI into team structures**

   Based on research on teamwork (Section 2.5) and on the teammate pattern (Section 4.2.1), it turns out that integrating AI into teams involves various levels of organizational effort. The lowest level occurs when implementing AI as a teammate is not intended. At level two, if using AI as a teammate in some selected teams is pursued, additional organizational practices are needed with respect to this dimension. The role AI might play within task sharing must be established, comparable to onboarding. Furthermore, continuous awareness of whether this role needs adjustment has to be practiced. This effort multiplies (level 3) if a company decides that most of its teams should systematically include AI as a supporting team member and should continuously negotiate on how AI can be optimally employed in this role. Thus, oversight and risk mitigation need to be included in organizational practices. If AI itself can take over coordinative tasks and decision-making (level 4) about the areas where it should be employed, the number of organizational steps might be reduced, but the complexity of deliberate managerial decision-making increases (Section 3.3).

   With regard to HCAI, reflection and oversight become easier because they are the responsibility of a team rather than a single individual. Exercising oversight and directly applying interventions become part of collaborative teamwork, as recommended from a socio-technical perspective (Section 2.1). However, if integrating AI as team members is a company-wide goal, risk management and maintaining fairness and control over AI can become more difficult.

4. **Quality management**

   Quality management (QM) is often combined with process management (Section 2.7). The top-down Pattern and the bottom-up pattern provide examples of how QM can be organized in different ways (Sections 4.1.2 and 4.1.3). In Case A, different opinions became apparent: some interviewees assumed that QM is only relevant during initial implementation and testing, while others supposed that it is continuously necessary. If management decides to use only AI systems, they consider highly reliable, the main organizational effort involves implementing cycles of testing and potential adjustment at the beginning (level 1). If reliability cannot be assumed, the application AI might be rejected. However, if the organization decides that using AI is still advantageous—even if some errors or biases must be corrected from time to time—routines of quality assurance have to be enacted. This includes medium effort if only individual workstations are equipped with AI as a powerful tool (Section 4.2.2) (level 2). If AI is used for collaborative work, or if quality assurance is organized as a mutual task (Sections 4.1.3 and 4.1.4), continuous quality checks involve higher organizational effort (level 3). This increases to the highest level if company-wide or network-wide business processes have to be reconsidered with respect to the continuous quality assurance of AI outcomes (level 4). An additional challenge is balancing or calibrating trust: if mistrust is exaggerated, efficiency decreases; undertrust can result in risks and damages.

   In relation to HCAI, QM is not only relevant for avoiding product defects or inadequate services, but also for ensuring fairness and eliminating biases that could harm people. QM is an important task that can be assigned to people and is part of a holistic job description in accordance with human-centered work design (Section 2.1).

5. **Explanations**
   The pattern on explainability and accountability (Section 4.1.5) describes the necessary organizational measures to achieve explainability. Explanations are usually used to support quality assurance but can also be relevant as input for learning, for justifying decisions, or contributing to innovation and continuous improvement (Section 2.3). The lowest level of organizational effort occurs when the AI system is considered reliable for routine tasks, and explanations are only relevant during testing. If AI output refers to exceptional cases, as in the context of predictive maintenance, explanations are more relevant (level 2), and organizational practices must ensure that there is enough time to consider explanations and that staff are prepared and encouraged to use them (Section 4.1.5). More organizational effort is required (level 3) when the creation of explanations is organized as a collaborative task involving different human experts (Section 4.1.7), as well as additional phases of collaborative reflection. The highest level is reached if explanations need to be disseminated and documented throughout the entire company to support organizational learning or to justify decisions to internal personnel or external partners.
   HCAI is explicitly addressed because it requires transparency and explainability. Both contribute to the crucial goal of accountability. Even if the efforts required for transparency reduce efficiency, it is important to give people a sense of control and to support trust-building. The more collaborative explainability is pursued and combined with reflection, the more communication between employees is promoted (Section 2.1).

6. **Continuous improvement and evolution of AI**
   The need for continuous improvement was discussed in Cases A and B. The patterns of bottom-up improvement (Section 4.1.3), appropriation (Section 4.1.4), and organizational learning (Section 4.1.6) demonstrate the relevance of organizational practices. The zero level means that the company simply does not care about adapting or continuously improving AI or curating relevant data, e.g., if the AI-supported task is not part of its core business. At a medium level of organizational effort (level 2), the company can establish an organizational practice of collecting experience with AI to derive requirements for improvement. Eventually, these requirements are conveyed to the suppliers of AI solutions. A high degree of organizational effort (level 3) is needed if there are routines (Section 4.1.3) at certain points in an organization for internally customizing AI, e.g., by adding rules or documents or adapting thresholds (Section 4.1.4). The highest level covers organizational practices in which AI is employed throughout the company and all types of usage, and their interplay is continuously subject to improvement or innovation efforts.
   The more improvement and innovation belong to the organizational culture (Section 2.2) of a company, the more interesting tasks are available for people, and the more opportunities arise for human-centered joint optimization in accordance with HCAI.

7. **Discussing and reflecting on AI experience in communicational processes**
   The pattern of integrating AI based tasks into communicational exchange (Section 4.1.7) characterizes certain types of organizational practices that demonstrate how AI usage can lead to more communication. If it is possible to restrict dealing with AI outcomes to individual workspaces of single employees, the need for communication about AI will be greatly reduced (level 1) and will mainly take place during the initial implementation. If employing AI is discussed from time to time with specialists or at meetings (level 2), organizational practices will need to provide opportunities that allow for this kind of communication. If the organization introduces systematic, regular exchange and reflection about AI results and their explanations, this requires additional effort (level 3). This level also applies if sporadic exchanges about AI cover the whole company. If systematic routines of communicational clarification about AI usage are introduced for the entire company and for interaction with partners, this indicates the highest effort (level 4). For all three higher levels, the organizational effort increases when communication content is additionally documented and exchanged via knowledge management systems to systematically support organizational learning.
   Communicational discussion and reflection are HCAI enablers. They help develop an organizational culture where oversight and overriding AI outcomes or experimenting with adaptations of AI are encouraged. Communication deepens explanations and allows for feedback that gives employees an impression of how well they interact with AI and supports their self-efficacy.

8. **Continuous Learning**
   The patterns derived from the empirical cases, particularly those concerning appropriation, organizational learning, and the sparring partner (Sections 4.1.4, 4.1.6, and 4.2.3), point toward the possibility of continuous learning on the job. If continuous learning of employees while using AI is not pursued by the organization, only marginal organizational practices are needed to ensure the initial learning before regular AI-usage (level 1). A medium level of effort (level 2) is needed to ensure occasional AI-related learning for individual professionals, e.g., with AI as a sparring partner (Section 4.2.3). A higher level of organizational practice takes place as a first step toward organizational learning (Section 4.1.6) if learning on the job is continuously integrated into collaborative problem solving, e.g., by teams. A very high effort (level 4) is required for AI-triggered company-wide learning that includes the exchange with partners and is integrated with communicational processes and knowledge management, which is also supported by AI (Section 2.3).

Continuous learning is crucial for the HCAI principle of human-guided co-learning and co-evolution (Xu & Gao, 2025). It is also a fundamental basis for the joint optimization of the interplay between humans and AI and is indispensable for the ongoing development of people's strengths.

With regard to the eight dimensions the question arises as to whether they are independent or could be partially merged. Apparently, there are partial overlaps. For example, discussion and reflection throughout communicational processes can be essential support for quality assurance, coordination of supportive task sharing in teams, or organizational learning. The tendency is that the more organizational effort is involved—as indicated in the third and fourth columns of Table 1—the more overlaps between the dimensions become relevant. However, each dimension can also serve its own purposes besides being a potential contributor to the other dimensions. For example, intensive communication can mainly satisfy human needs for social contact (Section 2.1). Communication, especially on the informal side, helps employees to see how others are working with AI. Additionally, it influences motivation and reflection on self-efficacy. Integration of AI into technical infrastructure or into team structures can, but need not, be interdependent. Furthermore, organization-wide learning in the course of AI usage can focus on domain expertise instead of dealing with AI in conjunction with the other dimensions.

To keep the AI-induced extent of organizational practices low, one can hypothesize that the more AI is implemented separately at an individual workstation without mutual dependence on cooperation with others, and the less quality management and continuous improvement are required, the lower the organizational effort. This statement is a kind of hypothesis that is not empirically substantiated but is analytically derived from the considerations above. It can promote further research and help practitioners make decisions between different options.

Similar hypotheses can be formed by listing the factors that drive the expansion of organizational practices: The organizational effort increases the more

a) unfinished, imprecise technology is used,
b) AI supports work on exceptional, innovative, or wicked problems where quality assurance can hardly be formalized,
c) AI is integrated into existing structures (technology, processes, team structures),
d) working on AI supported tasks is continuously combined with appropriation and customization to achieve quality assurance and improvement,
e) working on AI-supported tasks is continuously combined with learning to support personal development,
f) a transition takes place from AI usage for individual task handling to the support of collaborative work,
g) a transition is pursued from the simple options of accepting or rejecting AI results to frequently adjusting details of AI outcomes, or even to contributing to the continuous improvement of the employed AI systems.

These factors—that require more effort in implementing organizational practices—are the same ones that also promote the relevance of including human strength and development when dealing with AI. Fostering this relevance of human contributions is a basis of HCAI and emphasizes that humans are in the lead to ensure quality assurance and fairness as well as continuous improvement and innovation.

−

# 6. Conclusion

In what follows, we explain how the previous sections answer the initial questions (Section 1.2).

1. **Which existing approaches that address the relation of organizational design and information technology are relevant for HCAI?**
   The answer is given in Section 2 by describing a series of approaches that are relevant when implementing and using AI. They demonstrate possible roles and activities of human staff and the relevance of autonomy, flexibility, collaboration, and learning for both individuals and the meso-level of organizations. The subsection about work design (Section 2.1) offers criteria from the viewpoint of work psychology and socio-technical design that correspond to the goals of HCAI. It is outlined that management decisions need to be transformed into conventions and organizational practices, and that this transformation requires supportive communicational exchange. Systematic communication is also relevant for organizational learning, which contributes to the HCAI principle of co-learning and co-evolution. Organizational learning requires knowledge management, which will be complemented by AI. Management strategies must be considered with respect to HCAI, particularly to support human autonomy, e.g., via management by delegation. A central socio-technical issue is to consider teamwork as a unit of design. HCAI will be influenced by the question of how AI is integrated into teams. Furthermore, organizational approaches are relevant that consider technology not only on the level of workplaces but also with respect to larger units such as networks, communities or business processes. These are based on technical infrastructure that includes various components and that will become integrated with AI.

2. **What are different ways of organizationally embedding the usage of AI and how do these ways influence the goals of HCAI?**
Section 3.3 explains the details of keeping the organization in the loop. These details have been derived from the case studies. They cover, on the one hand, eight descriptions of highly interactive organizational practices. The description refers to the purposes of these practices, which are in accordance with HCAI principles. Most important is the coordination of how AI can be used in a way that allows users oversight, consideration and reflection on explanations, as well as flexible adaptation of AI outcomes. This flexibility needs to be interwoven with the co-evolution of human and AI. The use of AI must be aligned with preparing the staff and with the task to be carried out. The contextual factors and the situation of partners require continuous awareness. On the other hand, five types of managerial activities are detailed that are relevant for initiating the organizational practices.
This systematic consideration in Section 3.3 does not mirror the variety of possible ways in which the organizational practices can be realized. This variety of possibilities is described through ten patterns that deal with different ways of integrating AI into business processes, organizational learning, or communication processes. Learning, as a basis for HCAI-related evolution on the human side, can be coupled with the appropriation of AI or with using it as a sparring partner. Another crucial aspect for HCAI is the way oversight is organized—bottom-up or top-down. Organizational practices can accompany certain roles of AI, such as team member versus powerful tool. Every pattern is discussed with regard to its effect on HCAI. The more a pattern accepts the unreliability, incompleteness, and continuous evolution of AI, the more the relevance of human strengths is acknowledged and employed. According to the patterns, humans can contribute to quality assurance and fairness as well as to continuous improvement, if their experience and ability for mutual support, teamwork, informal interaction and learning on the job remain involved.

3. **Which dimensions help to understand the organizational effort implied by various patterns of embedding AI, and what are the HCAI-related drivers of investing this effort?**
By only understanding the ten patterns and their impact, one does not know the obstacles or enablers that influence how far management will implement the concept of keeping the organization in the loop. Thus, Section 5 identifies eight dimensions for various levels, considering what effort is needed if AI is integrated with organizational practices. The dimensions are derived from and aligned with the patterns of Section 4. They cover the integration with business processes, technical infrastructure, team structure, quality management, considering and reflecting explanations, continuous improvement, communicational processes, and continuous learning. Section 5 also outlines drivers for the expansion of organizational practices. Regarding HCAI, they imply the need to include human actors, enable communication and learning as part of personal development, emphasize the relevance of oversight and reflection, and highlight the need for allocating complete and significant tasks to the staff (Herrmann et al., 2018). A high level of integration will result in
- overall improvement and innovation of AI functionality, and companies' overall performance,
- an organizational culture that maximizes the benefit of using AI and developing the staff together,
- new roles and new tasks that emerge to keep people in the loop.

The consideration of the dimensions and drivers in Section 5 allows for the hypothesis that the higher the level of integration effort, the higher the compliance with HCAI. However, the effort that needs to be invested in the organizational practices of extensive integration can also turn out to be a barrier, since it involves informal structures and more people who might pursue diverging interests and can thus counteract top-down supervision and control.
The eight dimensions systematically help researchers and practitioners empirically investigate how mature companies are in pursuing the concept of keeping the organization in the loop. They are therefore also heuristics for assessing commitment to HCAI.

Reflecting on these results reveals that the expansion and characteristics of organizational practices are confronted with some essential contradictions: while imperfectness might cause damage, inefficiency, and stress for the staff, it also emphasizes the role of humans since they have to deal with the deficits of AI and help to overcome them. While centralized top-down control might help to check the staff's reliability in handling AI and whether trust is appropriately calibrated, it might neglect possibilities for innovation and personal development. Bottom-up strategies have the potential to enable people to participate in continuous improvement with beneficial effects on the evolution of the whole company, including productivity gains.

Being aware of these contradictions, it becomes clear that there is no single best selection of patterns and levels of integration that fits every company or organizational network. Each organization needs to decide—in relation to its specific situation and context—between waiting until AI is highly reliable for certain purposes or trying to benefit from imperfect solutions. Compromises might be needed that combine the imperfectness of AI with tight monitoring and oversight, informal responsiveness, continuous improvement, and innovation through organizational learning. For example, early adopters can benefit from AI before others do if they find ways of dealing with the challenges of imperfectness. It has to be taken into account that AI is a technology that could help to deal with its own shortcomings by supporting human-centered organizational practices that pursue the compensation of AI's weaknesses. Furthermore, AI can also be considered a kind of technology that contributes autonomously to its own evolution. This does not necessarily imply that AI adapts itself, but it can also contribute to continuous improvement by making proposals that are assessed and eventually enacted by human agents. Overall, AI is not only a subject of socio-technical design but can also contribute to it.

Finally, essential shortcomings have to be mentioned: the patterns as well as the dimensions of Table 1 are not empirically derived or substantiated. Due to the fast development of AI-related innovations that have to be addressed by organizational practices, it is not the time to wait until substantial empirical research is available. Dealing with the issue of uncertainty implies that solutions and decisions on how to organize AI implementation and usage effectively are very context-sensitive and differ from organization to organization, especially if human needs have to be concretely respected.

What is needed—but not dealt with in this chapter—are methods to measure the success of AI and the degree of achieving HCAI goals, again possibly by using AI. These methods will help to base the continuous evolution and adjustment of organizational practices on feedback regarding their success.

# 7. References

Acharya, D. B., Kuppan, K., & Divya, B. (2025). Agentic AI: Autonomous Intelligence for Complex Goals—A Comprehensive Survey. *IEEE Access*, *13*, 18912–18936. https://doi.org/10.1109/ACCESS.2025.3532853

Alavi, M., & Leidner, D. E. (2001). Knowledge management and knowledge management systems: Conceptual foundations and research issues. *MIS Quarterly*, 107–136.

Alter, S. (2002). The Work System Method for Understanding Information Systems and Information Systems Research. *Communications of the Association for Information Systems*, *9*, 90–104.

Argote, L., Lee, S., & Park, J. (2021). Organizational Learning Processes and Outcomes: Major Findings and Future Research Directions. *Management Science*, *67*(9), 5399–5429. https://doi.org/10.1287/mnsc.2020.3693

Aubouin-Bonnaventure, J., Fouquereau, E., Coillot, H., Lahiani, F. J., & Chevalier, S. (2021). Virtuous Organizational Practices: A New Construct and a New Inventory. *Frontiers in Psychology*, *12*. https://doi.org/10.3389/fpsyg.2021.724956

Bandura, A. (1977). Self-efficacy: Toward a unifying theory of behavioral change. *Psychological Review*, *84*(2), 191–215. https://doi.org/10.1037/0033-295X.84.2.191

Barbado, A., Corcho, Ó., & Benjamins, R. (2022). Rule extraction in unsupervised anomaly detection for model explainability: Application to OneClass SVM. *Expert Systems with Applications*, *189*, 116100. https://doi.org/10.1016/j.eswa.2021.116100

Berretta, S., Tausch, A., Ontrup, G., Gilles, B., Peifer, C., & Kluge, A. (2023). Defining human-AI teaming the human-centered way: A scoping review and network analysis. *Frontiers in Artificial Intelligence*, *6*, 1250725. https://doi.org/10.3389/frai.2023.1250725

Boud, D. (1985). *Reflection: Turning experience into learning*. Kogan Page.

By, R. T. (2005). Organisational change management: A critical review. *Journal of Change Management*, *5*(4), 369–380. https://doi.org/10.1080/14697010500359250

Cai, C. J., Winter, S., Steiner, D., Wilcox, L., & Terry, M. (2019). "Hello AI": Uncovering the Onboarding Needs of Medical Practitioners for Human-AI Collaborative Decision-Making. *Proceedings of the ACM on Human-Computer Interaction*, *3*(CSCW), 1–24. https://doi.org/10.1145/3359206

Castañer, X., & Oliveira, N. (2020). Collaboration, Coordination, and Cooperation Among Organizations: Establishing the Distinctive Meanings of These Terms Through a Systematic Literature Review. *Journal of Management*, *46*(6), 965–1001. https://doi.org/10.1177/0149206320901565

Cheng, M., Luo, Y., Ouyang, J., Liu, Q., Liu, H., Li, L., Yu, S., Zhang, B., Cao, J., Ma, J., Wang, D., & Chen, E. (2025). *A Survey on Knowledge-Oriented Retrieval-Augmented Generation* (arXiv:2503.10677). arXiv. https://doi.org/10.48550/arXiv.2503.10677

Cherns, A. (1976). The principles of sociotechnical design. *Human Relations*, *29*(8), 783–792.

Cherns, A. (1987). Principles of Sociotechnical Design Revisted. *Human Relations*, *40*(3), 153–162.

Cooley, M. (1996). On Human-Machine Symbiosis. In K. S. Gill (Ed.), *Human Machine Symbiosis* (pp. 69–100). Springer London. https://doi.org/10.1007/978-1-4471-3247-9_2

Crootof, R., Kaminski, M. E., & Price II, W. N. (2023). Humans in the Loop. *76 Vanderbilt Law Review*, (76/2/429). https://scholarship.law.vanderbilt.edu/vlr/vol76/iss2/2

Dellermann, D., Calma, A., Lipusch, N., Weber, T., Weigel, S., & Ebel, P. (2019). The future of human-AI collaboration: A taxonomy of design knowledge for hybrid intelligence systems. *Proceedings of the 52nd Hawaii International Conference on System Sciences (HICSS)*.

Dubey, A., Abhinav, K., Jain, S., Arora, V., & Puttaveerana, A. (2020). HACO: A Framework for Developing Human-AI Teaming. *Proceedings of the 13th Innovations in Software Engineering Conference on Formerly Known as India Software Engineering Conference*, 1–9. https://doi.org/10.1145/3385032.3385044

Dumas, M., La Rosa, M., Mendling, J., & Reijers, H. A. (2013). *Fundamentals of Business Process Management*. Springer.

Ehsan, U., & Riedl, M. O. (2020). Human-Centered Explainable AI: Towards a Reflective Sociotechnical Approach. In C. Stephanidis, M. Kurosu, H. Degen, & L. Reinerman-Jones (Eds.), *HCI International 2020—Late Breaking Papers: Multimodality and Intelligence* (Vol. 12424, pp. 449–466). Springer International Publishing. https://doi.org/10.1007/978-3-030-60117-1_33

Fischer, G., & Herrmann, T. (2011). Socio-technical systems: A meta-design perspective. *International Journal of Sociotechnology and Knowledge Development (IJSKD)*, *3*(1), 1–33. https://doi.org/DOI:%252010.4018/jskd.2011010101

Fogliato, R., Chappidi, S., Lungren, M., Fitzke, M., Parkinson, M., Wilson, D., Fisher, P., Horvitz, E., Inkpen, K., & Nushi, B. (2022). *Who Goes First? Influences of Human-AI Workflow on Decision Making in Clinical Imaging* (arXiv:2205.09696). arXiv. http://arxiv.org/abs/2205.09696

Garcia, E., Costa, A., Palanca, J., Giret, A., Julian, V., & Botti, V. (2020). Requirements for an Intelligent Maintenance System for Industry 4.0. In T. Borangiu, D. Trentesaux, P. Leitão, A. Giret Boggino, & V. Botti (Eds.), *Service Oriented, Holonic and Multi-agent Manufacturing Systems for Industry of the Future* (pp. 340–351). Springer International Publishing. https://doi.org/10.1007/978-3-030-27477-1_26

Goedicke, M., & Herrmann, T. (2008). A Case for ViewPoints and Documents. In B. Paech & C. Martell (Eds.), *Innovations for Requirement Analysis. From Stakeholders' Needs to Formal Designs* (Vol. 5320, pp. 62–84). Springer Berlin Heidelberg. https://doi.org/10.1007/978-3-540-89778-1_8

Grote, G., Ryser, C., Wäler, T., Windischer, A., & Weik, S. (2000). KOMPASS: A method for complementary function allocation in automated work systems. *International Journal of Human-Computer Studies*, *52*(2), 267–287. https://doi.org/10.1006/ijhc.1999.0289

Gulati, R., Puranam, P., & Tushman, M. (2012). Meta-organization design: Rethinking design in interorganizational and community contexts. *Strategic Management Journal*, *33*(6), 571–586. https://doi.org/10.1002/smj.1975

Gutterman, A. (2024). Definitions and Models of Organizational Culture. *SSRN Electronic Journal*. https://doi.org/10.2139/ssrn.4967434

Hackman, J. R., Oldham, G., Janson, R., & Purdy, K. (1975). A New Strategy for Job Enrichment. *California Management Review*, *17*(4), 57–71. https://doi.org/10.2307/41164610

Hackman, J. R., & Oldham, G. R. (1980). *Work redesign* (Vol. 279). Prentice Hall.

Hammer, M. (1990). Reengineering Work: Don't Automate, Obliterate. *Harvard Business Review*, *68*(4), 104–112.

He, V. F., Von Krogh, G., & Sirén, C. (2022). Expertise Diversity, Informal Leadership Hierarchy, and Team Knowledge Creation: A study of pharmaceutical research collaborations. *Organization Studies*, *43*(6), 907–930. https://doi.org/10.1177/01708406211026114

Herrmann, T. (2009). Systems Design with the Socio-Technical Walkthrough: In B. Whitworth & A. De Moor (Eds.), *Handbook of Research on Socio-Technical Design and Social Networking Systems* (pp. 336–351). IGI Global. https://doi.org/10.4018/978-1-60566-264-0.ch023

Herrmann, T. (2020). Socio-Technical Design of Hybrid Intelligence Systems–The Case of Predictive Maintenance. In H. Degen & S. Ntoa (Eds.), *Artificial Intelligence in HCI: First International Conference, AI-HCI 2020, Held as Part of the 22nd HCI International Conference, HCII 2020, Copenhagen, Denmark, July 19–24, 2020, Proceedings 22* (pp. 298–309). Springer International Publishing. https://doi.org/10.1007/978-3-030-50334-5_20

Herrmann, T. (2022). Promoting Human Competences by Appropriate Modes of Interaction for Human-Centered-AI. In H. Degen & S. Ntoa (Eds.), *Artificial Intelligence in HCI* (Vol. 13336, pp. 35–50). Springer International Publishing. https://doi.org/10.1007/978-3-031-05643-7_3

Herrmann, T. (2023). Collaborative Appropriation of AI in the Context of Interacting with AI. In H. Degen & S. Ntoa (Eds.), *Artificial Intelligence in HCI* (Vol. 14051, pp. 249–260). Springer Nature Switzerland. https://doi.org/10.1007/978-3-031-35894-4_18

Herrmann, T. (2024). Comparing Socio-technical Design Principles with Guidelines for Human-Centered AI. In H. Degen & S. Ntoa (Eds.), *Artificial Intelligence in HCI* (Vol. 14735, pp. 60–74). Springer Nature Switzerland. https://doi.org/10.1007/978-3-031-60611-3_5

Herrmann, T. (2025a). AI as a Sparring Partner – An HCAI Approach to Promote Human Capabilities. In H. Degen & S. Ntoa (Eds.), *Artificial Intelligence in HCI* (pp. 162–176). Springer Nature Switzerland.

Herrmann, T. (2025b). Intervenability as a Design Requirement for Autonomy and Oversight within Human-Centered AI. In C. K. Coursaris, J. Beringer, P.-M. Léger, & B. Öz (Eds.), *The Design of Human-Centered Artificial Intelligence for the Workplace* (pp. 143–166). Springer Nature Switzerland. https://doi.org/10.1007/978-3-031-83512-4_9

Herrmann, T., Ackermann, M. S., Goggins, S. P., Stary, C., & Prilla, M. (2018). Designing Health Care That Works – Socio-technical Conclusions. In *Designing Healthcare That Works. A Socio-technical Approach.* (pp. 187–203). Academic Press.

Herrmann, T., & Hoffmann, M. (2005). The Metamorphoses of Workflow Projects in their Early Stages. *Computer Supported Cooperative Work*, *14*(5), 399–432.

Herrmann, T., Jahnke, I., & Nolte, A. (2022). A problem-based approach to the advancement of heuristics for socio-technical evaluation. *Behaviour & Information Technology*, *41*(14), 3087–3109. https://doi.org/10.1080/0144929X.2021.1972157

Herrmann, T., & Loser, K.-U. (1999). Vagueness in models of socio-technical systems. *Behavior & Information Technology: Special Issue on Analysis of Cooperation and Communication*, *18*(5), 313–323.

Herrmann, T., & Nolte, A. (2024). Promoting the Adoption of AI-Based Recommendations Through Organizational Practices. In R. Agrifoglio, A. Lazazzara, & S. Za (Eds.), *Navigating Digital Transformation* (Vol. 73, pp. 195–212). Springer Nature Switzerland. https://doi.org/10.1007/978-3-031-76970-2_13

Herrmann, T., & Pfeiffer, S. (2023a). *Keeping the organization in the loop: A socio-technical extension of human-centered artificial intelligence*. *38*, 1523–1542. https://doi.org/10.1007/s00146-022-01391-5

Herrmann, T., & Pfeiffer, S. (2023b). Keeping the Organization in the Loop as a General Concept for Human-Centered AI: The Example of Medical Imaging. *Proceedings of the 56th Hawaii International Conference on System Sciences (HICSS)*, 5272–5281.

Holbeche, L. (2019). Designing sustainably agile and resilient organizations. *Systems Research and Behavioral Science*, *36*(5), 668–677. https://doi.org/10.1002/sres.2624

Husainah, N., & Ahmad, G. (2024). Delegation Leadership in Multinational Corporate Organizations. *Seascapeid Journal of Economics, Management, and Business*, *1*(2), 55–61.

Jarrahi, M. H., Askay, D., Eshraghi, A., & Smith, P. (2023). Artificial intelligence and knowledge management: A partnership between human and AI. *Business Horizons*, *66*(1), 87–99. https://doi.org/10.1016/j.bushor.2022.03.002

Kienle, A., & Herrmann, T. (2003). Integration of communication, coordination and learning material-a guide for the functionality of collaborative learning environments. *System Sciences, 2003. Proceedings of the 36th Annual Hawaii International Conference on System Sciences (HICSS)*, 10-pp.

Kirsch, C., Troxler, P., & Ulich, E. (1995). Integration of people, technology and organization: The european approach. In Y. Anzai, K. Ogawa, & H. Mori (Eds.), *Symbiosis of Human and Artifact* (Vol. 20, pp. 957–961). Elsevier. https://doi.org/10.1016/S0921-2647(06)80337-0

Kraut, R. E., Fish, R. S., Root, R. W., & Chalfonte, B. L. (1993). Informal Communication in Organizations: Form, Function, and Technology. In Baecker (Ed.), *Readings in Groupware and computer-supported Cooperative Work* (pp. 145–199). Morgan Kaufman. http://www.sociotech-lit.de/KFRC93-ICi.pdf

Kreutzer, M., Cardinal, L. B., Walter, J., & Lechner, C. (2016). Formal and Informal Control as Complement or Substitute? The Role of the Task Environment. *Strategy Science*, *1*(4), 235–255. https://doi.org/10.1287/stsc.2016.0019

Kroll, B., Schaffranek, D., Schriegel, S., & Niggemann, O. (2014). System modeling based on machine learning for anomaly detection and predictive maintenance in industrial plants. *Proceedings of the 2014 IEEE Emerging Technology and Factory Automation (ETFA)*, 1–7.

Levitt, J. (2003). *Complete Guide to Preventive and Predictive Maintenance*. Industrial Press Inc.

Luhmann, N. (1995). *Social Systems*. Stanford University Press.

Maciariello, J. A. (2016). Management by Objectives and Self-control. In M. Augier & D. J. Teece (Eds.), *The Palgrave Encyclopedia of Strategic Management* (pp. 1–4). Palgrave Macmillan UK. https://doi.org/10.1057/978-1-349-94848-2_165-1

Mahale, Y., Kolhar, S., & More, A. S. (2025). A comprehensive review on artificial intelligence driven predictive maintenance in vehicles: Technologies, challenges and future research directions. *Discover Applied Sciences*, *7*(4). https://doi.org/10.1007/s42452-025-06681-3

Mark, G. (2002). Conventions for Coordinating Electronic Distributed Work: A Longitudinal Study of Groupware Use. *Distributed Work*, 259–282.

Mathebula, B., & Barnard, B. (2020). The factors of delegation success: Accountability, compliance and work quality. *Expert Journal of Business and Management*, *8*(1), 76–97.

Mende, H., Peters, A., Ibrahim, F., & Schmitt, R. H. (2022). Integrating deep learning and rule-based systems into a smart devices decision support system for visual inspection in production. *Procedia CIRP*, *109*, 305–310. https://doi.org/10.1016/j.procir.2022.05.254

Mohrman, S. A., Cohen, S. G., & Mohrman, A. M. (1995). *Designing team-based organizations: New forms for knowledge work* (1st ed). Jossey-Bass.

Moran, J. W., & Brightman, B. K. (2000). Leading organizational change. *Journal of Workplace Learning*, *12*(2), 66–74. https://doi.org/10.1108/13665620010316226

Muller, M., & Weisz, J. (2022). Extending a Human-AI Collaboration Framework with Dynamism and Sociality. *Proceedings of the 1st Annual Meeting of the Symposium on Human-Computer Interaction for Work*, 1–12. https://doi.org/10.1145/3533406.3533407

Mumford, E. (1983). *Designing Human Systems for New Technology: The ETHICS Method*. Manchester Business School. https://books.google.de/books?id=JTjxIwAACAAJ

Mumford, E. (1995). *Effective systems design and requirements analysis: The ETHICS approach*. Macmillan.

Mumford, E. (2000). A Socio-Technical Approach to Systems Design. *Requirements Engineering*, *5*, 125–133.

Okamura, K., & Yamada, S. (2020). Adaptive trust calibration for human-AI collaboration. *PLOS ONE*, *15*(2), e0229132. https://doi.org/10.1371/journal.pone.0229132

O'Neill, T., McNeese, N., Barron, A., & Schelble, B. (2022). Human–Autonomy Teaming: A Review and Analysis of the Empirical Literature. *Human Factors: The Journal of the Human Factors and Ergonomics Society*, *64*(5), 904–938. https://doi.org/10.1177/0018720820960865

Orlikowski, W. J. (1996). Improvising organizational transformation over time: A situated change perspective. *Information Systems Research*, *7*(1), 63–92.

Phillips, N., & Lawrence, T. B. (2012). The turn to work in organization and management theory: Some implications for strategic organization. *Strategic Organization*, *10*(3), 223–230. https://doi.org/10.1177/1476127012453109

Pipek, V. (2005). *From tailoring to appropriation support: Negotiating groupware usage*. Oulun Yliopisto.

Piplica, D., Peronja, I., & Luković, T. (2024). Controlling in the Function of Management by Objectives. *DIEM: Dubrovnik International Economic Meeting*, *9*(1), 131–142.

Powell, W. W. (1990). Neither Market nor Hierarchy. Network Forms of Organization. *Research in Organizational Behavior*, *12*, 295–336.

Rahwan, I. (2018). Society-in-the-loop: Programming the algorithmic social contract. *Ethics and Information Technology*, *20*(1), 5–14.

R'bigui, H., & Cho, C. (2017). The state-of-the-art of business process mining challenges. *International Journal of Business Process Integration and Management*, *8*(4), 285. https://doi.org/10.1504/IJBPIM.2017.088819

Reijers, H., & Mansar, S. L. (2005). Best practices in business process redesign: An overview and qualitative evaluation of successful redesign heuristics. *Omega*, *33*(4), 283–306. https://doi.org/10.1016/j.omega.2004.04.012

Saastamoinen, H., & White, G. M. (1995). On handling exceptions. *Proceedings of Conference on Organizational Computing Systems - COCS '95*, 302–310. https://doi.org/10.1145/224019.224051

Schein, E. H. (1990). Organizational culture. *American Psychologist*, *45*(2), 109–119. https://doi.org/10.1037/0003-066X.45.2.109

Shneiderman, B. (2020). Human-Centered Artificial Intelligence: Three Fresh Ideas. *AIS Transactions on Human-Computer Interaction*, 109–124. https://doi.org/10.17705/1thci.00131

Shneiderman, B. (2022). *Human-Centered AI*. Oxford University Press.

Stravinskiene, I., & Serafinas, D. (2020). The Link between Business Process Management and Quality Management. *Journal of Risk and Financial Management*, *13*(10), 225. https://doi.org/10.3390/jrfm13100225

Tadesse Bogale, A., & Debela, K. L. (2024). Organizational culture: A systematic review. *Cogent Business & Management*, *11*(1). https://doi.org/10.1080/23311975.2024.2340129

Timofeev, A. V., & Denisov, V. M. (2020). Machine Learning Based Predictive Maintenance of Infrastructure Facilities in the Cryolithozone. In E. Pricop, J. Fattahi, N. Dutta, & M. Ibrahim (Eds.), *Recent Developments on Industrial Control Systems Resilience* (pp. 49–74). Springer International Publishing. https://doi.org/10.1007/978-3-030-31328-9_3

Trist, E. (1981). The Evolution of Socio-Technical Systems. In A. H. Van de Ven & W. F. Joyce (Eds.), *Perspectives on organization design and behavior*. Wiley New York.

Trist, E., & Bamforth, K. (1951). Some social and psychological consequences of the long wall method of coal getting. *Human Relations*, *4*, 3–38.
Verbeke, W. (2000). A revision of Hofstede et al.'s (1990) organizational practices scale. *Journal of Organizational Behavior*, *21*(5), 587–602. https://doi.org/10.1002/1099-1379(200008)21:5%253C587::AID-JOB22%253E3.0.CO;2-5
Wenger, E. (1999). *Communities of practice: Learning, meaning, and identity*. Cambridge University Press.
Winby, S., & Xu, W. (2026). Human-Centered AI Maturity Model (HCAI-MM): An Organizational Design Perspective. In W. Xu (Ed.), *Handbook of Human-Centered Artificial Intelligence*. Springer Nature.
Wright, C. (1993). Taylorism Reconsidered: The Impact of Scientific Management within the Australian Workplace. *Labour History*, (64), 34. https://doi.org/10.2307/27509164
Xu, W., & Gao, Z. (2024). Applying HCAI in Developing Effective Human-AI Teaming: A Perspective from Human-AI Joint Cognitive Systems. *Interactions*, *31*(1), 32–37. https://doi.org/10.1145/3635116
Xu, W., & Gao, Z. (2025). An Intelligent Sociotechnical Systems (iSTS) Framework: Enabling a Hierarchical Human-Centered AI (hHCAI) Approach. *IEEE Transactions on Technology and Society*, *6*(1), 31–46. https://doi.org/10.1109/TTS.2024.3486254
Zhang, R., McNeese, N. J., Freeman, G., & Musick, G. (2021). "An Ideal Human": Expectations of AI Teammates in Human-AI Teaming. *Proceedings of the ACM on Human-Computer Interaction*, *4*(CSCW3), 1–25. https://doi.org/10.1145/3432945
Zonta, T., da Costa, C. A., da Rosa Righi, R., de Lima, M. J., da Trindade, E. S., & Li, G. P. (2020). Predictive maintenance in the Industry 4.0: A systematic literature review. *Computers & Industrial Engineering*, *150*, 106889. https://doi.org/10.1016/j.cie.2020.106889